%% file: kskl_prc.tex
\documentclass[%
  reprint,
superscriptaddress,
 amsmath,amssymb,
 aps,
]{revtex4-2}

\usepackage{graphicx}
\usepackage{dcolumn}
\usepackage{bm}
\usepackage{hyperref}
\usepackage[normalem]{ulem}
\usepackage{xcolor} 
\usepackage{orcidlink}

\begin{document}


\title{Measurement of Spin-Density Matrix Elements in $\phi(1020)\to K_S^0K_L^0$ Photoproduction with a Linearly Polarized Photon Beam at $E_\gamma=8.2-8.8$~GeV}

\include{authors.tex}



\date{\today}

\begin{abstract}
We measure the spin-density matrix elements (SDMEs) for the photoproduction of $\phi(1020)$ off of the proton in its decay to $K_S^0K_L^0$, using 105~pb$^{-1}$ of data collected with a linearly polarized photon beam using the GlueX experiment.
The SDMEs are measured in nine bins of the squared four-momentum transfer $t$ in the range $-t=0.15-1.0$~GeV$^2$, providing the first measurement of their $t$-dependence for photon beam energies $E_\gamma = 8.2-8.8$~GeV.
We confirm the dominance of Pomeron exchange in this region, and put constraints on the contribution of other Regge exchanges.
We also find that helicity amplitudes where the helicity of the photon and the $\phi(1020)$ differ by two units are negligible.
\end{abstract}

\maketitle


\section{Introduction}

The study of the bound states of Quantum Chromodynamics provides valuable insight into the theory's non-perturbative regime and phenomena such as quark confinement.  Hadrons containing strange quarks are intriguing since their masses lie between hadrons composed only of the light up and down quarks and those containing heavy charm and bottom quarks. The strange-quark hadrons can therefore shed light on the transition from relativistic bound states made of light quarks to non-relativistic bound states containing a heavy quark. In particular, the spectrum of $s\bar{s}$ strangeonium mesons remains less well understood than that of mesons composed primarily of up and down quarks~\cite{deQuadros:2020ntn}. With the advent of high-precision experiments targeting the spectroscopy of light- and strange-quark mesons with masses up to 2.5~GeV, we anticipate significant progress in this area. 
One less studied process for the production of strangeonium is in photoproduction, which is now accessible at the GlueX experiment.
A natural starting point for studying the photoproduction of strangeonium is the lightest predominantly $s\bar{s}$ meson, the $\phi(1020)$.

For incident beam energies $E_\gamma = 8.2-8.8$~GeV, meson production off of a proton target primarily proceeds through diffractive production.  
The angular dependence of vector meson photoproduction can be described by relating the spin-density matrices of the incoming photon $\rho(\gamma)$ and the produced vector meson $\rho(V)$ through the production amplitude $T$~\cite{schilling}:
\begin{equation}
    \rho(V) = T\rho(\gamma)T^\dagger.
    \label{eqn:prod}
\end{equation}
Eleven of these spin-density matrix elements (SDMEs) can be measured in photoproduction experiments, and nine SDMEs can be measured using a linearly polarized photon beam~\cite{schilling}. 
At sufficiently large beam energies, only two of these nine SDMEs are expected to be non-zero in the helicity frame, due to $s$-channel helicity conservation (SCHC)~\cite{schc1,schc2,schc3}.  
Determining the applicability of SCHC for $\phi(1020)$ photoproduction
 at $E_\gamma = 8.2-8.8$~GeV will help in studying excited $\phi$ mesons in photoproduction.

The Joint Physics Analysis Center (JPAC) recently developed a model that describes the diffractive photoproduction of light vector mesons with a polarized photon beam using Regge-theory amplitudes~\cite{Mathieu:2018xyc}.  
This model uses Regge theory amplitudes fitted to pre-2020 measurements, and predicts $\phi(1020)$ photoproduction at $E_\gamma = 8.5$~GeV to proceed primarily via Pomeron exchange with a small contribution from $\pi$ and $\eta$ exchange.  
The Pomeron is a helicity-preserving exchange particle with natural parity.  The $\pi$ and $\eta$ are unnatural-parity exchange particles.
The JPAC model was recently shown to describe GlueX measurements of $\rho(770)$ SDMEs well~\cite{gluex-rho}.

Most measurements of polarized $\phi(1020)$ photoproduction have been at photon beam energies $E_\gamma < 3$~GeV.
The LEPS Collaboration measured $\phi(1020)$ SDMEs and cross sections for a linearly polarized photon beam with $E_\gamma = 1.5 - 2.9$~GeV \cite{leps2005,leps2010,leps2017}.  
The CLAS Collaboration measured cross sections and SDMEs with an unpolarized photon beam with $E_\gamma = 2.0 - 2.8$~GeV~\cite{clas2014}, and with a linearly polarized photon beam with $E_\gamma = 1.5 - 2.1$~GeV~\cite{clasthesis} in an unpublished Ph.D. thesis.  
The general conclusion from these analyses is that SCHC does not hold at these beam energies, and that production mechanisms beyond Pomeron exchange are required.

The few existing measurements of polarized $\phi(1020)$ photoproduction with $E_\gamma > 3$~GeV have low statistical precision.
Measurements of vector-meson photoproduction were performed at SLAC in 1973 using a linearly polarized photon beam with $E_\gamma=2.8$, 4.7, and 9.3~GeV~\cite{vector-slac}.  
While the total and differential cross sections were measured at each of the three energies, due to the limited statistical precision of the data, the lowest two energy bins were combined and only the $\phi(1020) \to K^+K^-$ decay mode was considered when measuring the SDMEs, yielding 53 events for $E_\gamma=2.8$ and 4.7~GeV, and 61 events for $E_\gamma=9.3$~GeV.  
Events corresponding to $\phi(1020) \to K_S^0K_L^0$ were also identified by this measurement, but only about $5-10$ in each energy bin, which was considered insufficient to measure SDMEs.
In 1985, the Omega Photon Collaboration measured cross sections 
and SDMEs in the $E_\gamma=20-40$~GeV range using 1135 $\phi(1020)\to K^+K^-$ decays~\cite{phi-omega}.  
Both measurements were consistent with SCHC with natural parity exchange within their large uncertainties.
Our measurement of the $\phi(1020)$ SDMEs has sufficient statistical precision to study the production mechanisms at these higher energies in detail.

In this paper, we describe the measurement of $\phi(1020)$ SDMEs using a linearly polarized beam of $E_\gamma = 8.2-8.8$~GeV with the GlueX detector.  We analyze the reaction,
\begin{equation}
\gamma p \to \phi(1020) p,\quad \phi(1020) \to K_S^0 K_L^0, \quad K_S^0 \to \pi^+\pi^-,
\end{equation}
where the $K_L^0$ is not detected.  
The analyzed data correspond to a total integrated luminosity of 105~pb$^{-1}$.
This large sample of photoproduced $\phi(1020)$ allows us to measure the SDMEs in nine bins of Mandelstam-$t$ in the range $-t=0.15-1.0$~GeV$^2$.  
In Section~II we describe the experimental setup used to collect this data.
In Section~III we describe how we select events corresponding to the above reaction, while Section~IV discusses the model and fit procedure used to measure the SDMEs. 
The analysis procedure closely follows that of our recent publication on $\rho(770)$ SDMEs~\cite{gluex-rho}.
The results are presented and discussed in Section V.

\section{Experimental Setup and Simulations}

The GlueX experiment consists of a tagged photon beam and a large-acceptance spectrometer, and has been described previously in detail~\cite{gluex-nim,gluex-det1,gluex-det2,gluex-det3,gluex-det4,gluex-det5,gluex-det6,gluex-det7}.
The 12 GeV electron beam from the CEBAF accelerator is delivered in bunches separated by 4~ns, and is converted into a linearly polarized photon beam through coherent bremsstrahlung off of a 50~$\mu$m thick diamond radiator.  
The energy and time of the scattered electrons are measured in a dipole spectrometer. 
The photon beam travels 75~m to the main experimental hall, where it is collimated and its flux and polarization are measured, with an average degree of linear polarization of $P_\gamma\approx 35\%$ in the peak photon flux region used in this analysis.  
The photon beam is directed onto a 30~cm long liquid hydrogen target positioned in the middle of a 2T superconducting solenoid. The target is surrounded by a start counter, central and forward drift chambers, and a barrel calorimeter.  
A forward calorimeter and a time-of-flight wall downstream of the solenoid provide additional coverage in the forward direction. Charged and neutral particles with polar angles from $1^\circ$ to $150^\circ$ are detected.

We use samples of Monte Carlo simulated events to study the detector response and to determine the experimental acceptance.  The simulated $\gamma p \to \phi(1020) p \to  K_S^0K_L^0p$ events are generated with a $t$ distribution proportional to $e^{-b t}$ with a slope parameter of $b=4.4$~GeV$^{-2}$, and a $M(K_S K_L)$ distribution that follows a $P$-wave relativistic Breit-Wigner with parameters $M=1020$~MeV/$c^2$ and $\Gamma=4.2$~MeV. The $\phi(1020)\to K_S^0K_L^0$ decay angles are generated isotropically.  The simulated events are passed through a Geant4-based simulation~\cite{geant4} of the detector response and analyzed using the same procedures as the experimental data. 

\begin{figure}[!tb]
    \includegraphics[width=0.9\columnwidth]{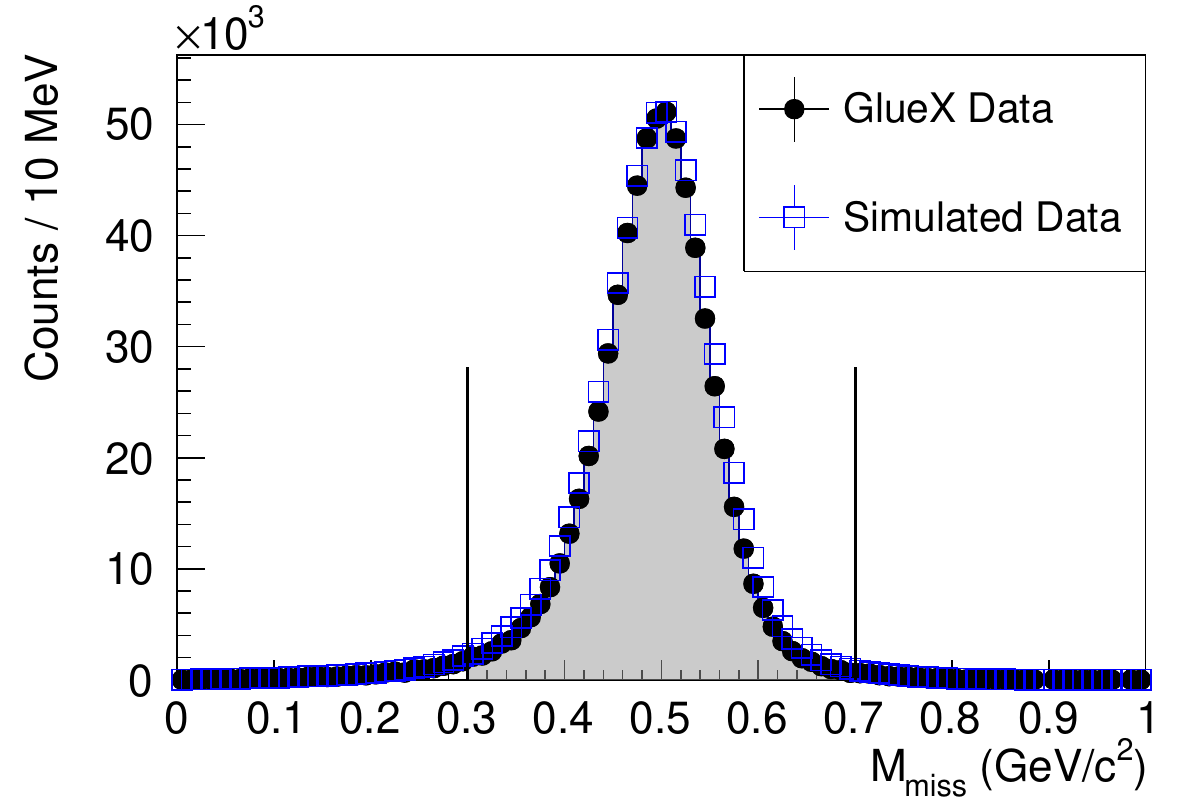}
    \includegraphics[width=0.9\columnwidth]{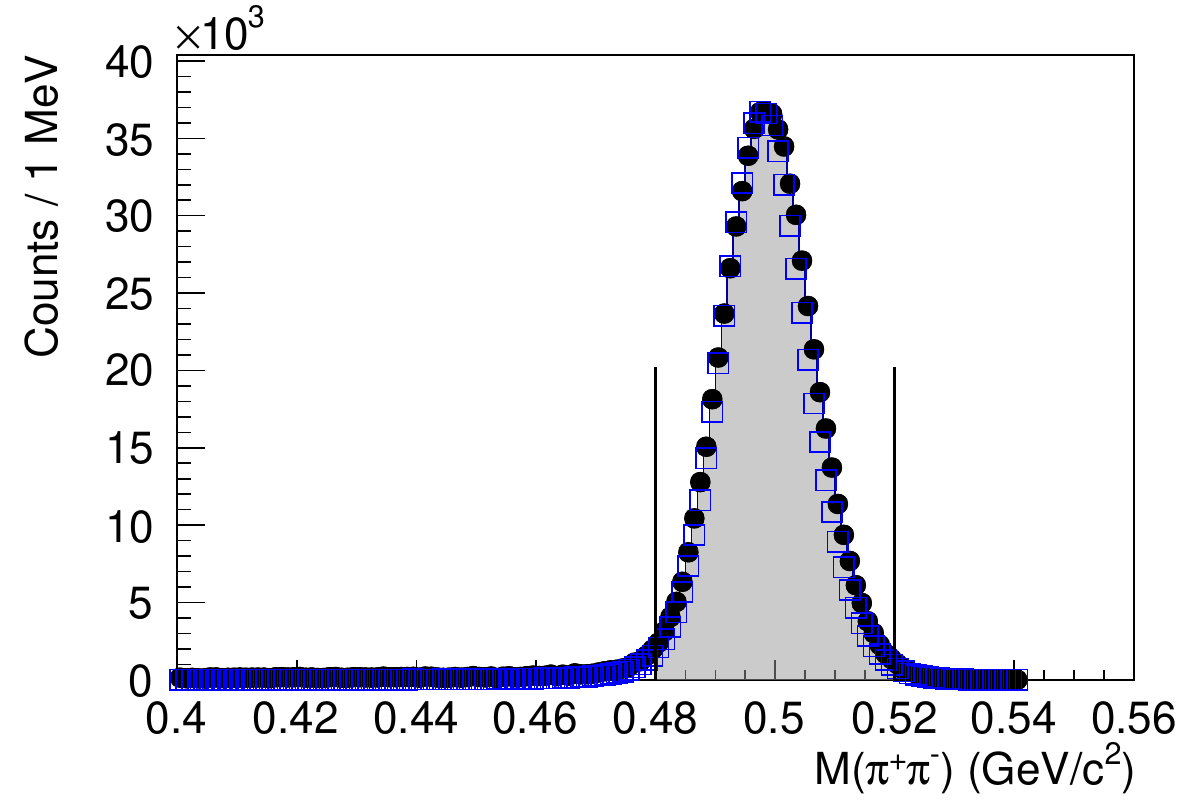}
    \caption{Distributions for selected $\gamma p \to K_S^0 K_L^0 p$ events where the $K_L^0$ is reconstructed as a missing particle, as described in the text: (top) the missing mass, showing a clear $K_L^0$ signal, and (bottom) the $\pi^+\pi^-$ invariant mass, showing a clear $K_S^0\to\pi^+\pi^-$ signal. The distributions for measured events (solid points) agree well with the simulated data (open squares).  The vertical lines show the regions selected for further analysis.}

    \label{fig:masses}
\end{figure}

\section{Data Analysis}

We reconstruct the reaction $\gamma p \to \phi(1020) p$ with the decay $\phi(1020) \to K_S^0 K_L^0$ by reconstructing the $K_S^0$ decaying into $\pi^+\pi^-$ along with the recoil proton, and treating the $K_L^0$ as a missing particle.
We select events with exactly three charged particle candidates, where we require a $\pi^{+}$, $\pi^{-}$ and a proton to be identified by loose selections on particle time-of-flight and ionization energy deposited in the drift chambers.
To retain $K_S^0K_L^0$ events that have extra calorimeter showers due to effects such as splitoffs from hadronic interactions of charged particles inside the calorimeters, we allow for up to two calorimeter showers not matched to a charged particle track.    These selections efficiently suppress $K^0_S K^0_S$ events that are kinematically similar to our $K^0_S K^0_L$ events, where the second $K^0_S$ decays to either $\pi^+ \pi^-$ or $\pi^0 \pi^0$.
The primary vertex is defined by the position of closest approach of the proton candidate to the beam axis, and has a resolution of $\approx3$~mm for the events analyzed in the this paper. Events are selected only if this vertex position lies inside the target region and is at least two cm away from the upstream and downstream ends of the target cell.

The energy of the beam photon candidates is required to be in the range with a high degree of linear polarization, $E_\gamma = 8.2-8.8$~GeV.  In addition, we require the measured times of the beam photon candidates and final-state particles to be consistent with coming from the same electron bunch.  Due to inefficiencies in the photon tagger and the finite spectrometer resolution, final-state particles can be matched with an incorrect beam photon candidate.  To subtract the contributions from such combinatorial mismatches, we assign a weight of one to all signal events, and assign weights of $-1/4$ to combinations of beam photons and final-state  particles that are mismatched by two or three beam bunches.  Events that are mismatched by one beam bunch are not included in the analysis. 

Due to their long lifetime of $c\tau \approx 15$~m, most $K_L^0$ mesons decay outside the spectrometer.  Instead of detecting the $K_L^0$, we infer its existence by utilizing 4-momentum conservation and considering the missing mass in the event, with all other particles detected.  We define the missing 4-momentum of an event as
\begin{equation}
    p_\mathrm{miss} = p_\gamma + p_\mathrm{target} - (p_p + p_{\pi^+} + p_{\pi^-}),
\end{equation}
where $p_\gamma$ and $p_\mathrm{target}$ are the 4-momenta of the beam photon candidate and the target proton, respectively, and $p_p$, $p_{\pi^+}$, and $p_{\pi^-}$ are the reconstructed 4-momenta of the final-state particles.  The distribution of the missing mass $M_\text{miss} = \sqrt{p_\text{miss}^2}$ is shown in Fig.~\ref{fig:masses}~(top) for events with $M(K_S^0K_L^0)< 1.1$~GeV after all other event selections are applied, where $M(K_S^0K_L^0)$ is calculated with the reconstructed $K_S^0\to\pi^+\pi^-$ 4-momentum and $p_\mathrm{miss}$ as the $K_L^0$ 4-momentum.  The missing mass distribution peaks near the nominal $K_L^0$ mass of 497.6~MeV~\cite{pdg} and the resolution is found to be well-modeled in simulation. We select events with $M_\text{miss} = 0.3-0.7$~GeV.

\begin{figure}[!tb]
    \includegraphics[width=0.9\columnwidth]{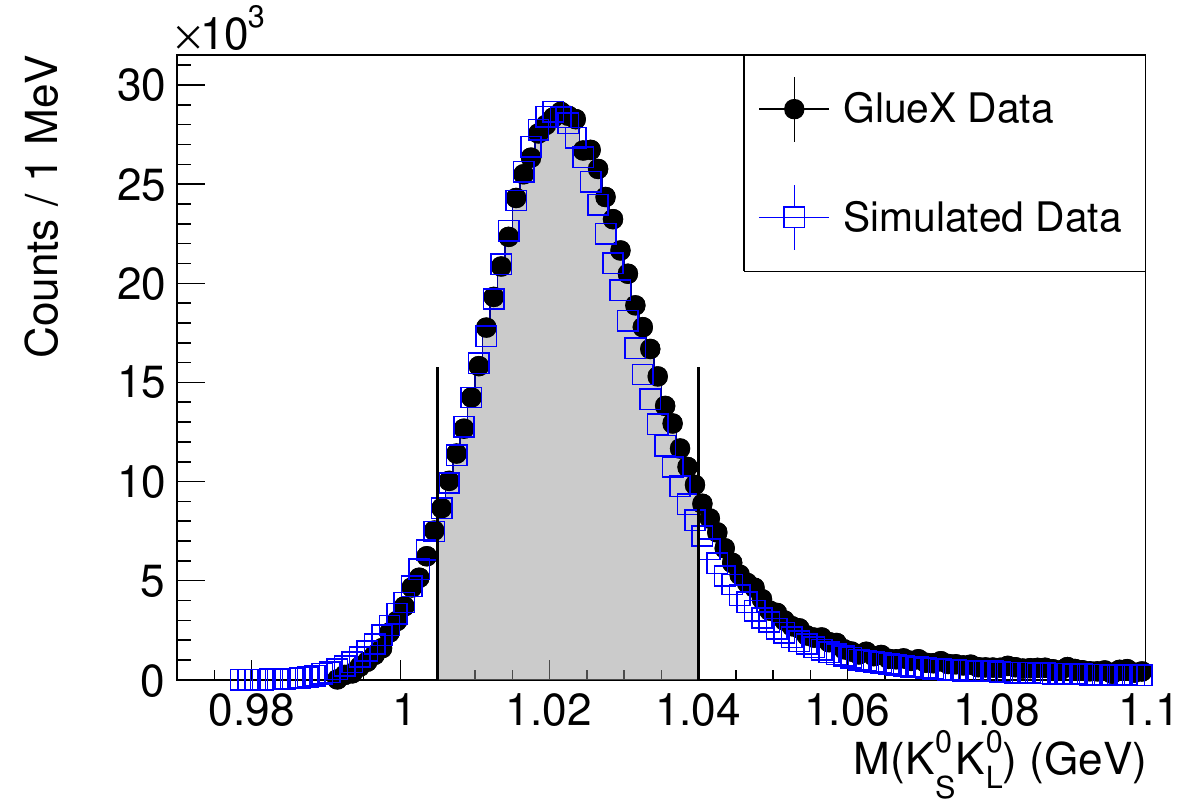}
    \includegraphics[width=0.9\columnwidth]{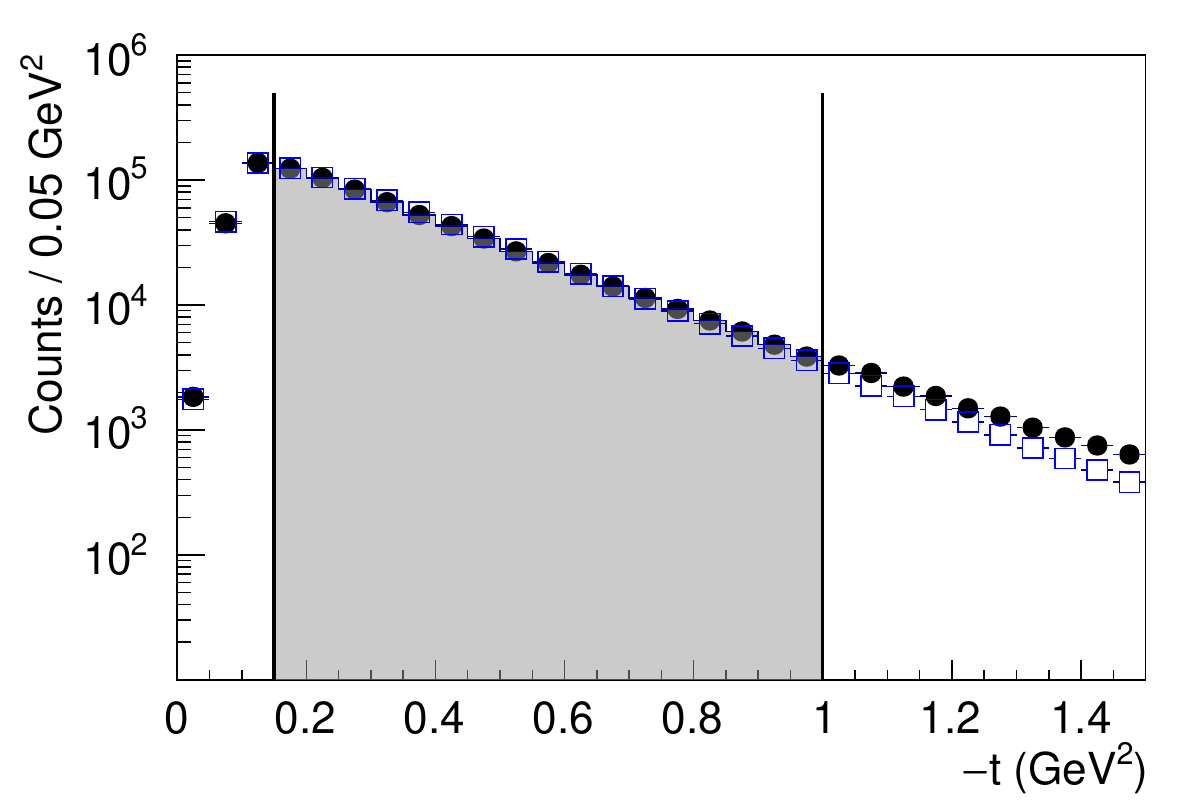}
    \caption{(Top) The $K_S^0K_L^0$ invariant mass distribution for selected events, showing a clear $\phi(1020)\to K_S^0K_L^0$ signal. 
    (Bottom) The Mandelstam-$t$ distribution for the events selected in the top panel.  The simulated events (blue squares) assume a constant exponential $t$-slope of $4.4~\mathrm{GeV^{-2}}$, and agree with the measured data (black points) up to $-t\approx1$~GeV$^2$. 
    The shaded regions indicate the events selected for the SDME analysis.
    The legend shown in the top figure applies to both.}

    \label{fig:ksklmassandt}
\end{figure}

To improve the resolution in the reconstruction of the $K_S^0\to\pi^+\pi^-$ decays and to suppress events with additional undetected particles beyond the $K_L^0$, we perform a kinematic fit on all events.  In these fits, we constrain $M_\text{miss}$ to the known $K_L^0$~\cite{pdg}, and constrain the $\pi^+\pi^-$ candidates to originate from a common secondary vertex which may be displaced from the primary reaction vertex defined by the $p$ candidate.  We select well-reconstructed events that satisfy this hypothesis by requiring $\chi^2/d.o.f. < 4$.  
All event distributions and fits described in this paper use the 4-vectors resulting from this kinematic fit.

Events that do not contain a $K_S^0 \to \pi^+\pi^-$ candidate are suppressed by requiring the $K_S^0$ decay vertex to be displaced from the primary vertex.  
The $K_S^0$ mesons in this measurement are produced at small polar angles with momenta of several GeV, and therefore decay predominantly several centimeters from the primary vertex.  The charged pions from their decay are reconstructed in the forward drift chambers, with a vertex resolution of $\approx1$~cm, which leads to a good separation between primary and decay vertices.  
We calculate the $K_S^0$ flight significance, i.e. the magnitude of the displacement between the primary and decay vertices divided by the total uncertainty on this quantity, which is calculated from the position uncertainties of the primary and decay vertices.  We select events with $K_S^0$ flight significance $>4\sigma$, which is 74\% efficient for the $K_S^0\to\pi^+\pi^-$  decays in this measurement. 
Since the $\phi(1020)$ is the dominant feature in this low $K_S^0K_L^0$ mass region, the $\chi^2/d.o.f.$ and flight significance selections were chosen to maximize both the measured $\phi(1020)$ yield and the sample purity.  

The $\pi^+\pi^-$ invariant mass distribution after these event selections is shown in Fig.~\ref{fig:masses}~(bottom).  We observe a clear peak due to $K_S^0\to\pi^+\pi^-$ decays, with 98\% purity.  We select $K_S^0$ candidates by requiring $M(\pi^+\pi^-)=0.48-0.52$~GeV.  

We show the $K_S^0K_L^0$ invariant mass distribution after all event selections are applied in Fig.~\ref{fig:ksklmassandt}~(top), for the mass region near the $K\overline{K}$ threshold.  This spectrum is clearly dominated by the $\phi(1020)$.  We select events with $M(K_S^0K_L^0)=1.005-1.040$~GeV for the spin-density matrix element analysis. 

The distribution of the squared 4-momentum transfer $t = \left(p_\text{p} - p_\text{target}\right)^2$ is shown in Fig.~\ref{fig:ksklmassandt}~(bottom).  
As expected for a diffractive process, the distribution follows an exponential form and deviates from a single exponential slope only for $t\gtrsim 1$~GeV$^2$, beyond the range considered in this paper.
The drop in the distribution below $-t\lesssim 0.1$~GeV$^2$ reflects the limited acceptance for the reconstruction of the recoil proton in this region. To avoid this region, we require $-t >0.15$~GeV$^2$.

 After all event selections, the final sample contains approximately $6.5\times10^5$ $\phi(1020)$ mesons, representing a dataset four orders of magnitude larger than previous measurements at this beam energy.

\begin{figure}[!tb]
    \includegraphics[width=0.48\textwidth]{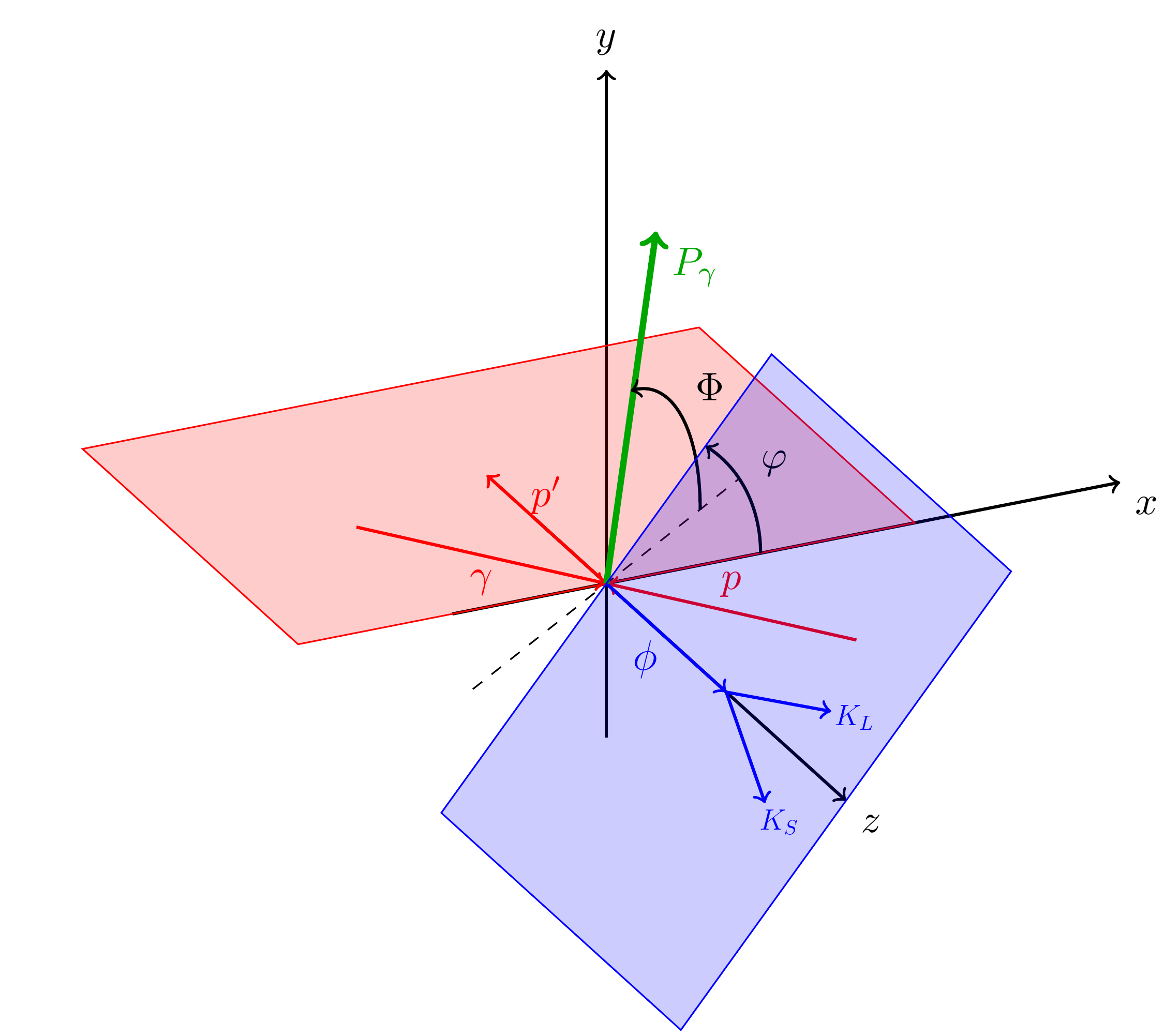}
    \caption{Definition of the angles in the center-of-mass frame for the decay $\phi(1020) \to K_S^0K_L^0$. The hadronic production plane is shown in red and the $\phi(1020)$ decay plane in blue. The polarization vector (green) has an angle $\Phi$ with respect to the hadronic production plane.}

    \label{fig:helicity_system}
\end{figure}

\section{Analysis Method}

To extract the spin-density matrix elements, we use the AmpTools framework~\cite{amptools} to perform an unbinned extended-maximum-likelihood fit of a reaction model to the measured events.  The formalism and method are described in detail in Ref.~\cite{gluex-rho}.  

Briefly, the number density $n(\vartheta,\varphi,\Phi)$ of vector mesons decaying into two spinless particles is proportional to a normalized angular distribution $W(\vartheta, \varphi, \Phi)$.  Here, $\vartheta$ and $\varphi$ are the polar and azimuthal angles of the decay particles defined in the helicity system of the vector meson, respectively, and $\Phi$ is the azimuthal angle between the beam photon polarization direction and the hadronic production plane in the center-of-mass frame of the reaction (see Fig.~\ref{fig:helicity_system}).
Defining the measured degree of linear polarization $P_\gamma$, the angular distribution is given by
\begin{widetext}
    \begin{align}
        W(\vartheta, \varphi, \Phi) & = W^0(\vartheta, \varphi) - P_\gamma \cos(2\Phi) \,  W^1(\vartheta, \varphi) - P_\gamma \sin(2\Phi) \, W^2(\vartheta, \varphi) \\
        \intertext{with}
        \nonumber W^0(\vartheta, \varphi) & = \frac{3}{4\pi} \left(\frac{1}{2}(1 - \rho_{00}^0) + \frac{1}{2}(3\rho_{00}^0 -1 )\cos^2\vartheta - \sqrt{2}\,\text{Re}\,\rho_{10}^0\,\sin2\vartheta \,\cos\hspace{0.5mm}\varphi - \rho_{1-1}^0 \sin^2\vartheta \, \cos2\varphi \right) \\
        \nonumber W^1(\vartheta, \varphi) & = \frac{3}{4\pi} \left( \rho_{11}^1\sin^2\vartheta + \rho_{00}^1\cos^2\vartheta - \sqrt{2}\,\text{Re}\,\rho_{10}^{1}\sin2\vartheta\, \cos\hspace{0.5mm}\varphi - \rho_{1-1}^1\sin^2\vartheta \cos2\hspace{0.5mm}\varphi \right) \\
        \nonumber W^2(\vartheta, \varphi) & = \frac{3}{4\pi} \left( \sqrt{2}\,\text{Im}\,\rho_{10}^2\sin2\vartheta \,\sin\hspace{0.5mm}\varphi + \text{Im}\,\rho_{1-1}^2\sin^2\vartheta \, \sin2\hspace{0.5mm}\varphi \right).
    \end{align}
    \label{eq:Wfunction}
\end{widetext}
Here, $W^0$ describes the unpolarized component of the angular distribution, while $W^1$ and $W^2$ describe the polarization-dependent components. These components are expressed in terms of the SDMEs $\rho^i_{jk}$, where $i=0,1,2$ and $j,k=-1,0,1$.

\begin{figure*}[!tb]
    \includegraphics[width=0.9\textwidth]{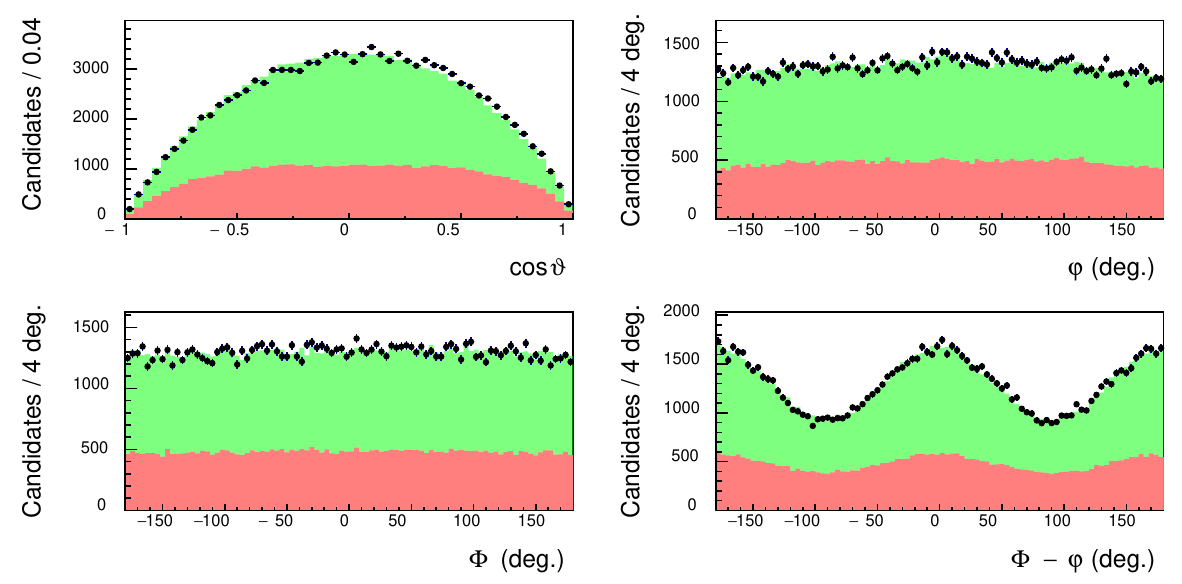}
    \caption{Comparison of measured angular distributions (black points) to accepted MC events weighted by the fit result (green) and the background from events with mismatched beam photons (red) for the range $-t=0.150-0.185$~GeV$^2$: (top left) the cosine of the helicity angle $\vartheta$, (top right) the helicity angle $\varphi$,  (bottom left) the azimuthal angle $\Phi$ of the beam photon polarization vector with respect to the production plane, (bottom right) the difference between $\Phi$ and $\varphi$.}

    \label{fig:fitdistributions}
\end{figure*}

\begin{figure*}[!tb]
    \includegraphics[width=0.3\textwidth]{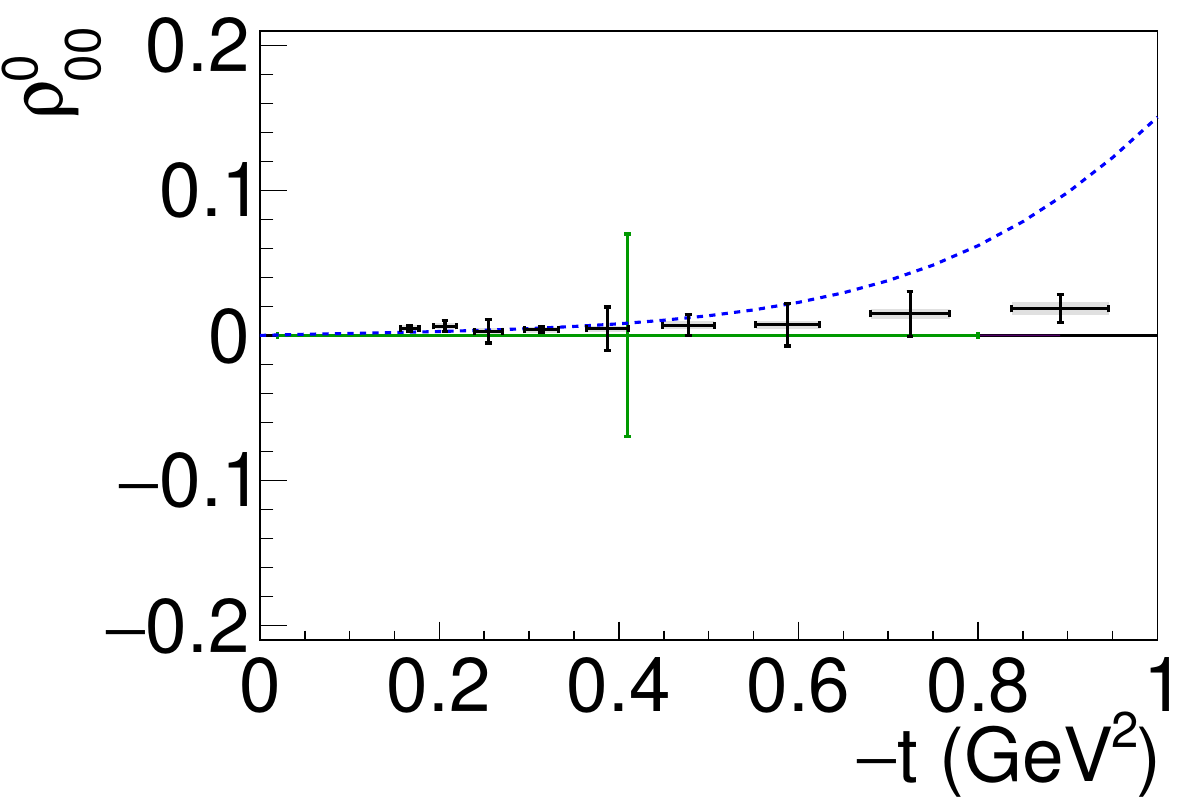}
    \includegraphics[width=0.3\textwidth]{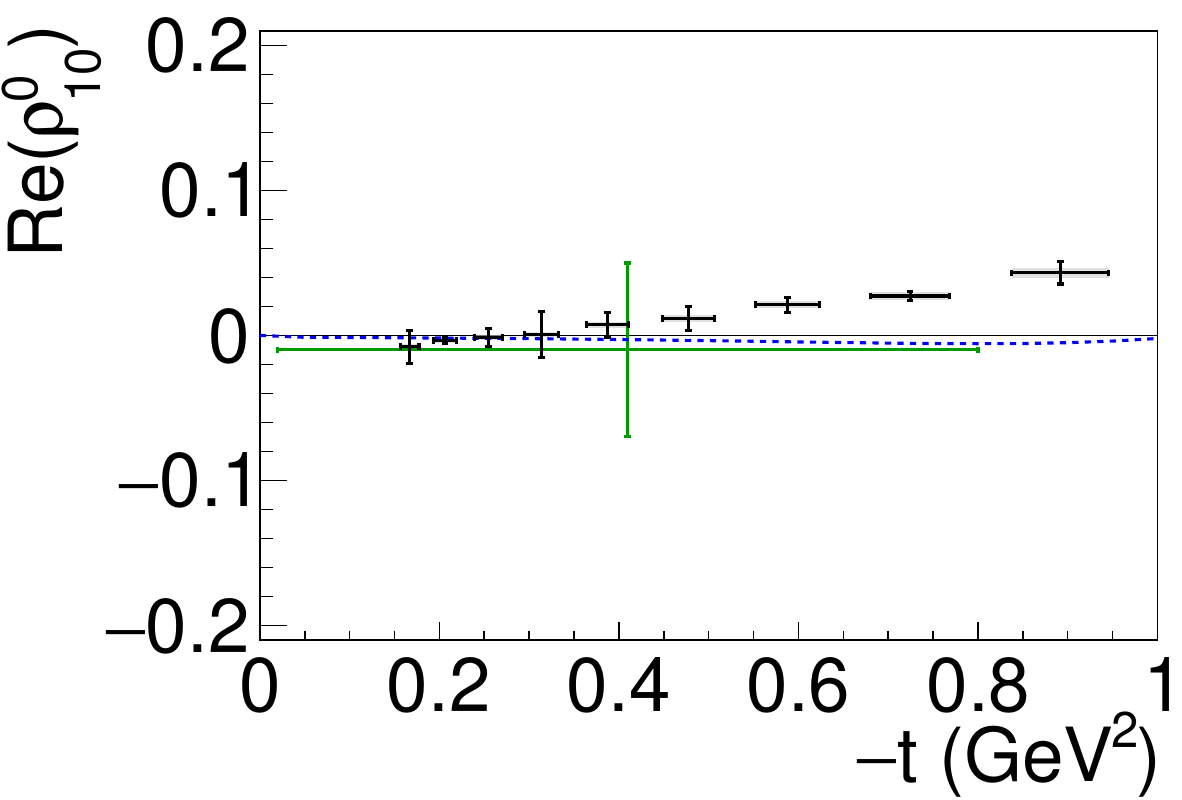}
    \includegraphics[width=0.3\textwidth]{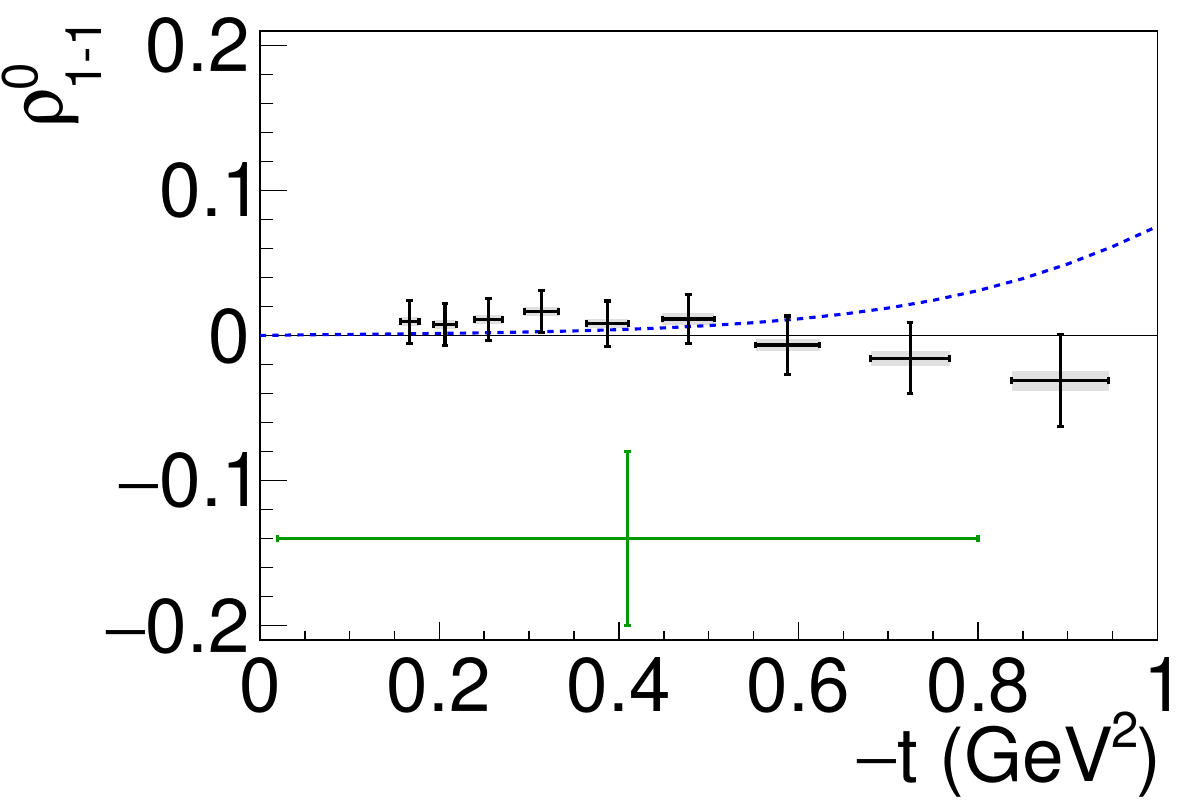}
    \includegraphics[width=0.3\textwidth]{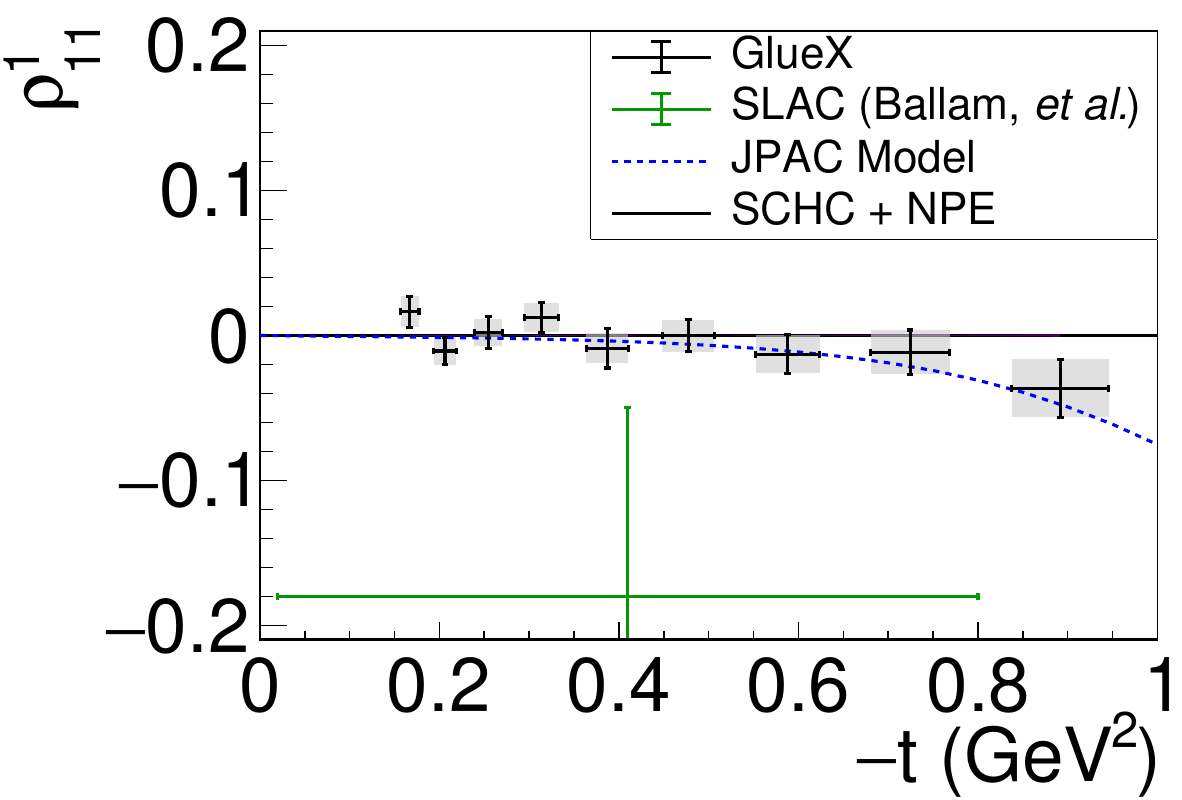}
    \includegraphics[width=0.3\textwidth]{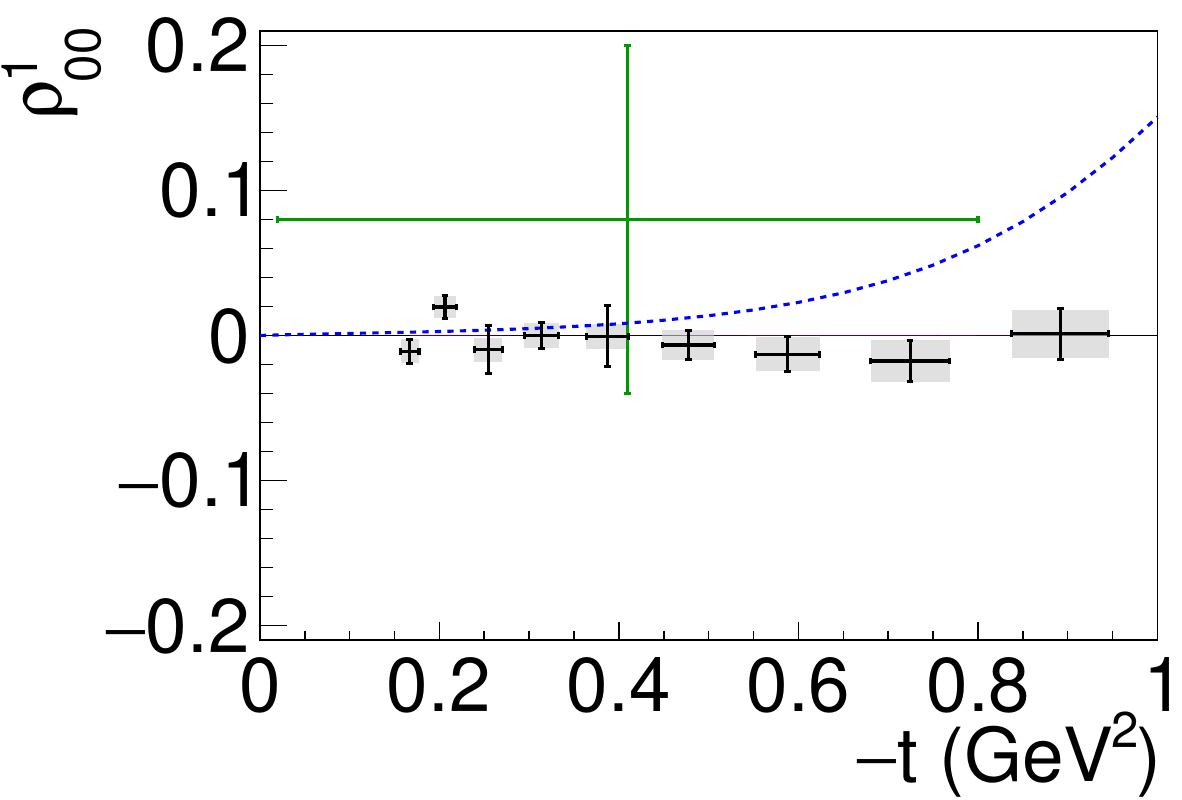}
    \includegraphics[width=0.3\textwidth]{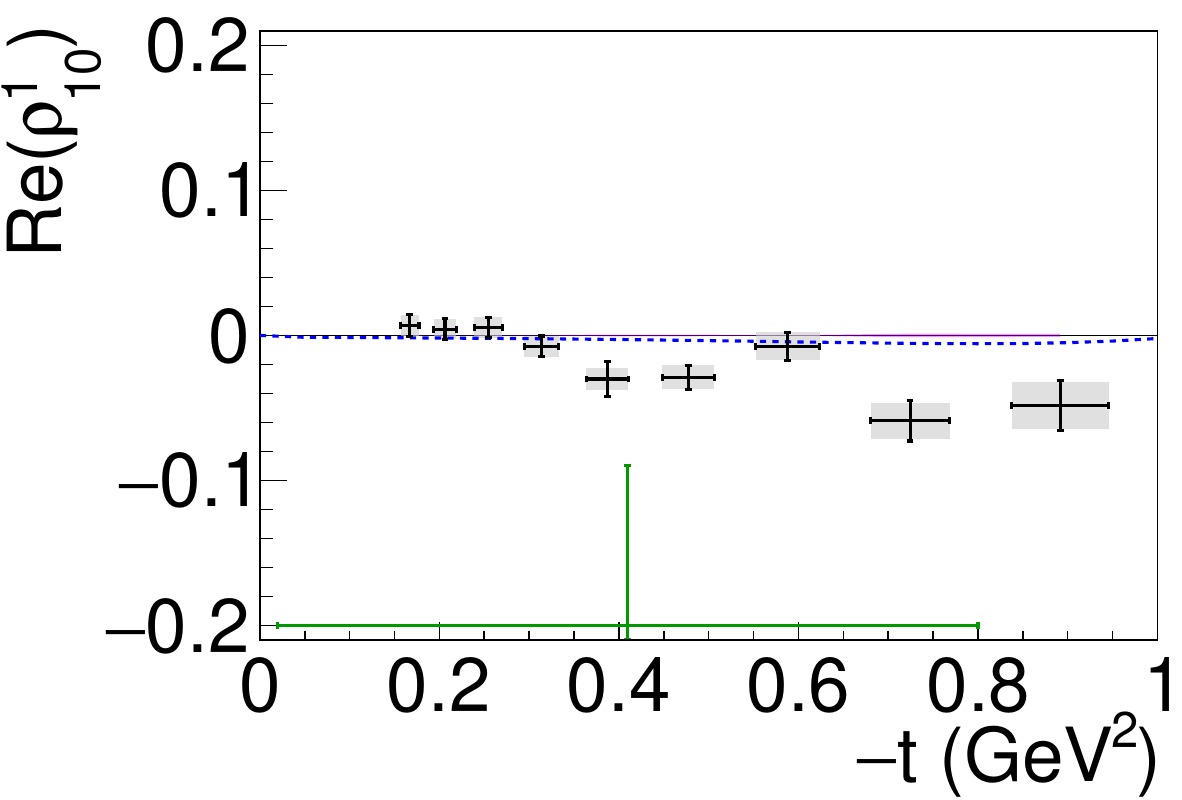}
    \includegraphics[width=0.3\textwidth]{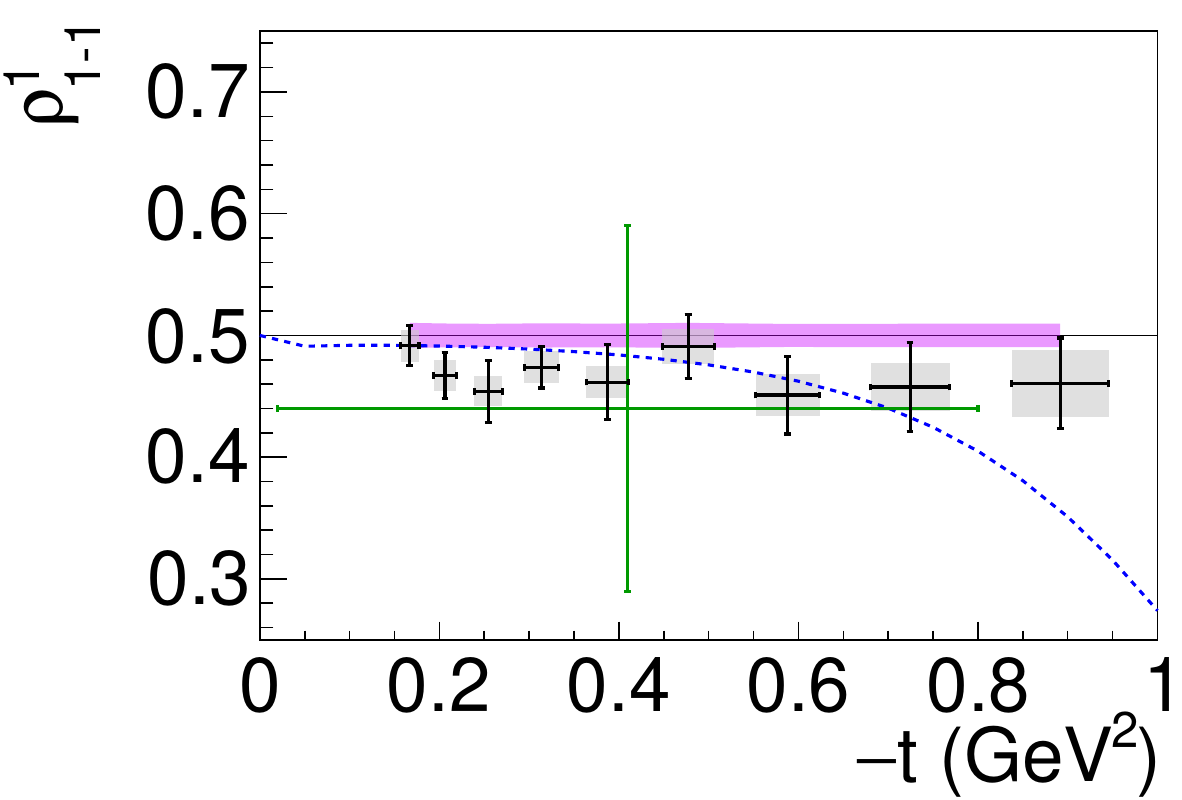}
    \includegraphics[width=0.3\textwidth]{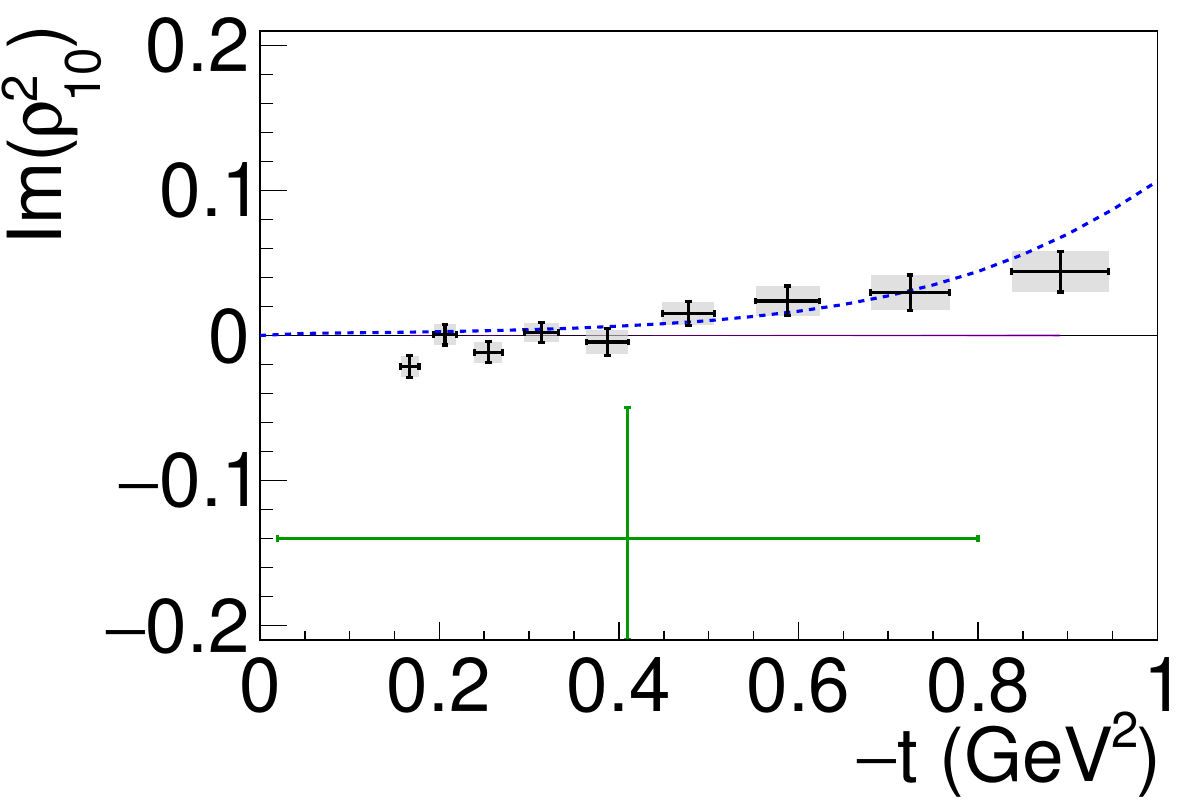}
    \includegraphics[width=0.3\textwidth]{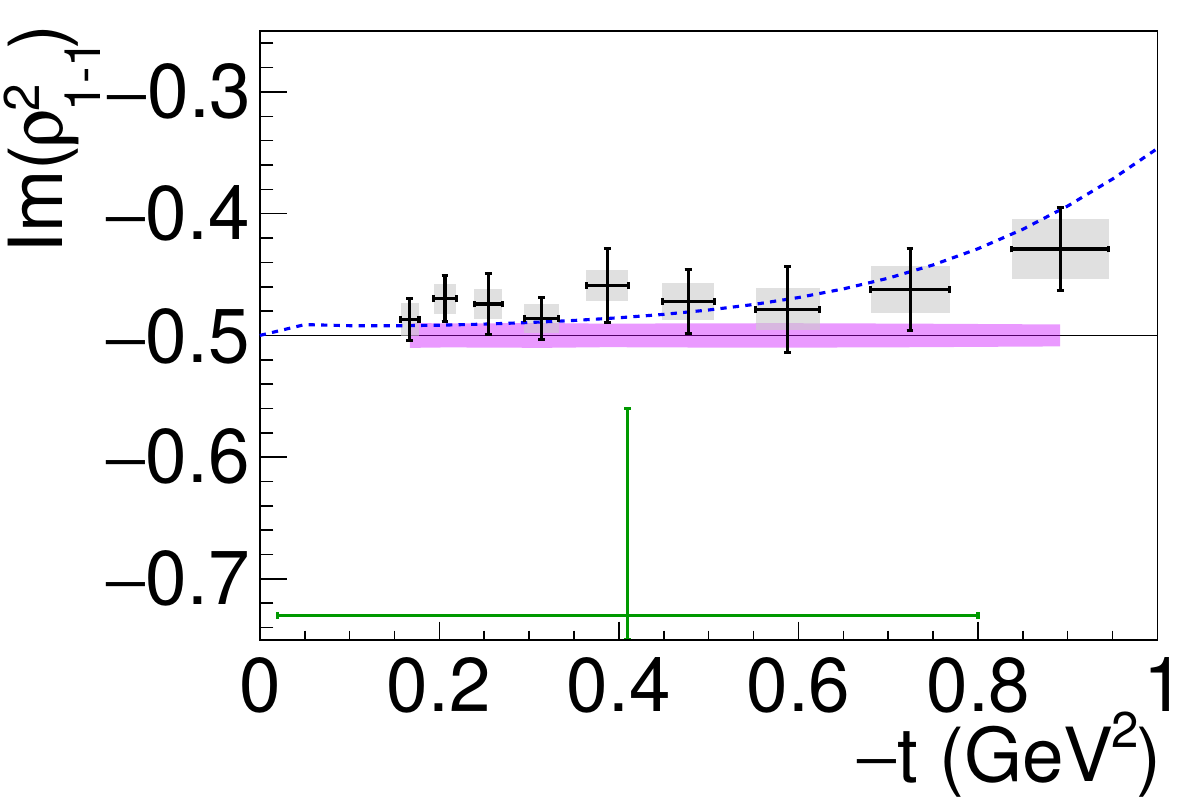}
    \caption{Spin-density matrix elements for $\phi(1020)$ mesons produced by a linearly polarized photon beam in the helicity frame.  Our measured values are represented by the black points, with shaded gray boxes indicating the statistical uncertainties, and the error bars represent the total uncertainties.  The correlated systematic uncertainties described in the text are shown as violet error bands. The measurements by Ballam \textit{et al.}~\cite{vector-slac}~(SLAC) are given by the green data points.  The horizontal solid lines show the expectation for $s$-channel helicity conservation with natural-parity exchange (SCHC + NPE), and the blue dashed lines show the Regge theory-based predictions of Ref.~\cite{Mathieu:2018xyc}.}

    \label{fig:results}
\end{figure*}

To obtain the SDMEs that best describe our measured events, we maximize the logarithm of the extended likelihood function
\begin{equation}
\mathcal{L} = 
\frac{e^{-(\bar{N}+\beta)}(\bar N + \beta)^{N}}{N!} \left(\prod^N_{i=1} n(\vartheta_i,\varphi_i,\Phi_i) \eta(\vartheta_i,\varphi_i,\Phi_i) \right) / \left( \prod^{\tilde{N}_B}_{i=1} n(\vartheta_i,\varphi_i,\Phi_i) \eta(\vartheta_i,\varphi_i,\Phi_i) \right)^{\frac{\beta}{\tilde{N}_B}}
\end{equation}
where $N$ is the total number of events, $\bar N$ is the expected number of signal events, $\beta$ is the estimated number of background events,  $\eta(\vartheta,\varphi,\Phi)$ is the experimental acceptance which is determined using the samples of Monte Carlo simulated events described previously, which reproduce the production kinematics but are isotropic in the decay angles, and $n(\vartheta_i,\varphi_i,\Phi_i)$ is the number density described in the previous paragraph.  
The background events in this case are due to incorrect matches between beam photons and final-state particles, and a separate data set with $\tilde{N}_B$ number of independent events, as described in Sec.~III, is used to estimate the number of accidental events $\beta$ in the signal data set.

We perform this fit independently in 9 bins of Mandelstam-$t$ in the range $-t = 0.15-1$~GeV$^2$.  The definition of the bins is given in Table~\ref{tab:sdme_values}.  We can then determine the $t$-dependence of the SDMEs from the results of these fits.

To evaluate the quality of each fit, we compare event distributions observed in the data to those from the accepted phase-space MC events which are weighted by $n(\vartheta, \varphi, \Phi)$ using maximum-likelihood estimates of the SDMEs.  As an example, we show a comparison in Fig.~\ref{fig:fitdistributions} for the range $-t=0.150-0.185$~GeV$^2$ for $\cos\vartheta$, $\varphi$, $\Phi$, and $\Phi - \varphi$ distributions.  In all four cases, the MC events weighted by the fit result show good agreement with the data.  A similar level of agreement is found in all nine bins of $-t$.

\section{Results}

Fig.~\ref{fig:results} shows the measured SDMEs as a function of Mandelstam-$t$.
The data points are positioned at the mean $t$ values of all events within each bin, while the horizontal error bars represent the standard deviation of the $t$ distribution in each bin.  
The vertical error bars represent the total uncertainty, which is the quadrature sum of the statistical and systematic uncertainties of each data point. The statistical uncertainties of each data point are represented by the shaded boxes. These statistical uncertainties are determined by the Bootstrap technique~\cite{bootstrap} based on fits to 500 resampled datasets.  
Correlated systematic uncertainties on the magnitude of the beam linear polarization and a correction factor for $\rho^0_{00}$, as described below, are shown by the shaded bands and are negligible, except for $\rho^1_{1-1}$ and $\mathrm{Im}(\rho^2_{1-1})$ SDMEs.  
Predictions from $s$-channel helicity conservation with natural-parity exchange (SCHC + NPE) and from the JPAC model~\cite{Mathieu:2018xyc} are also shown. At low $-t$, the measured SDMEs agree with the SCHC + NPE expectation, where only $\rho^1_{1-1}$ and $\mathrm{Im}(\rho^2_{1-1})$ are non-zero, with values of 0.5 and $-0.5$, respectively. At higher $-t$, the measured SDMEs deviate from SCHC + NPE, in line with the JPAC model predictions that attribute this to $\pi$ and $\eta$ exchange. Our data suggest a smaller contribution from this process than the model currently assumes.

Several sources of systematic uncertainties in these SDME measurements were investigated.
The largest systematic uncertainty on the polarized SDMEs comes from the measurement of the degree of linear beam polarization by a triplet polarimeter~\cite{gluex-det3}.
A total polarization uncertainty of 2.1\% is taken as a systematic uncertainty on the overall normalization of the polarized SDMEs $\rho^1_{ij}$ and $\rho^2_{ij}$.  
The orientations of the linear beam polarization in the lab frame are fixed parameters in our fits. However, a high-precision analysis of the decay asymmetry of $\gamma p \to \rho^0(770) p$ events with $\rho^0(770) \to \pi^+ \pi^-$ has revealed deviations from the nominal beam orientations by a few degrees.  In our fits, we fix the beam-polarization angles to the values estimated from the $\rho^0(770)$ decay asymmetry analysis, and determine the systematic uncertainty due to this choice by performing fits in which these angles were varied by $\pm1\sigma$ of their total uncertainty.  We take the largest of the deviations from the nominal from these two fits as the systematic uncertainty from this source for a given SDME.
The observed shifts of the SDME values were found to be small, but significant in many cases, particularly for $\rho^1_{ij}$.

To verify the analysis and fit procedure, we performed Monte Carlo studies using a sample of $2\times10^7$ $\phi(1020)\to K_S^0K_L^0$ events that were generated assuming SCHC and natural-parity exchange, and  that were analyzed in the same manner as the data.  All of the fitted SDMEs were found to be consistent with their generated values, except for $\rho^0_{00}$, whose fitted values were found to be consistent with an $t$-independent offset of $0.0075\pm0.0005$.  This offset was attributed to a bias in the measured direction of the $K_S^0$ 4-momentum arising from the separation between the primary vertex and the secondary $K_S^0\to\pi^+\pi^-$ vertex.  To correct for this bias, we correct the measured $\rho^0_{00}$ values by 0.0075 and assign a common systematic uncertainty of $\pm0.0005$. 

To investigate the impact of a small isotropic background underneath the $\phi(1020)$ meson peak, which could be due to an $S$-wave component of the reaction amplitude or some different incoherent source, we performed the SDME fits where we included a background term that was distributed isotropically in angular space and added incoherently to the model of Eq.~4.
The intensity of this background term varies from approximately 2\% to 5\% from the smallest to largest $-t$~bin.
Only $\rho^0_{00}$,  $\rho^1_{1-1}$, and $\mathrm{Im}(\rho^2_{1-1})$ were affected by this change, and the deviation from the nominal fit was taken as an additional systematic uncertainty due to the fit model for each point. 
For $\rho^1_{1-1}$ and $\mathrm{Im}(\rho^2_{1-1})$, the relative uncertainty associated with this systematic effect reach 7\%.  For $\rho^0_{0 0}$, it is the dominant source of uncertainty.

No systematic effect due to the event selection criteria was found to significantly affect the measured SDMEs.

\begin{figure*}[!tb]
    \includegraphics[width=0.48\columnwidth]{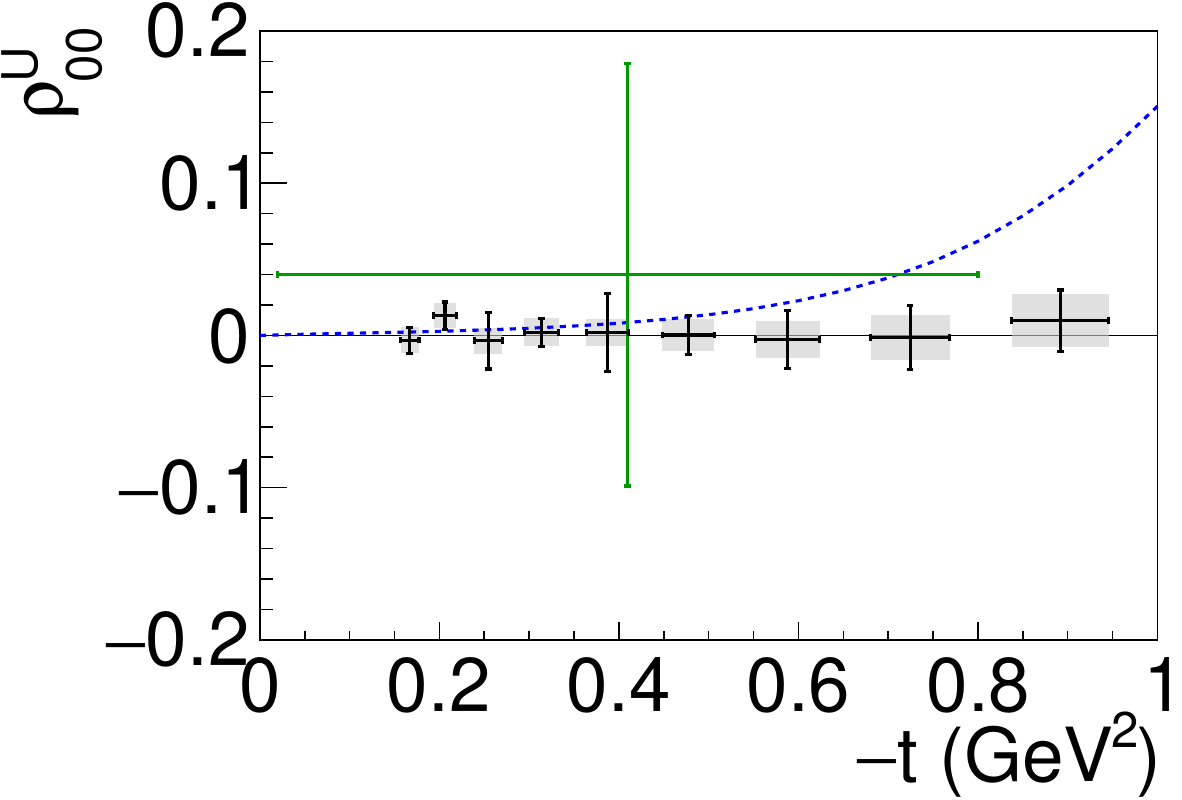}
    \includegraphics[width=0.48\columnwidth]{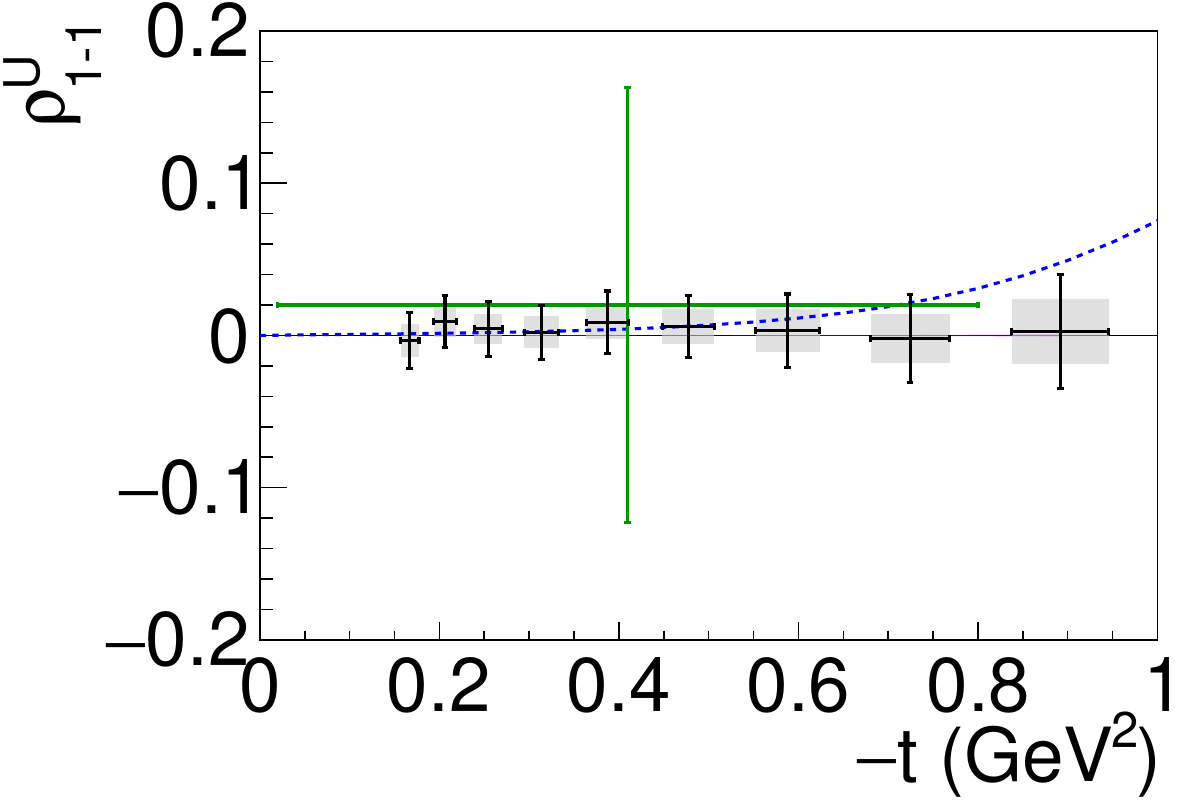}
    \includegraphics[width=0.48\columnwidth]{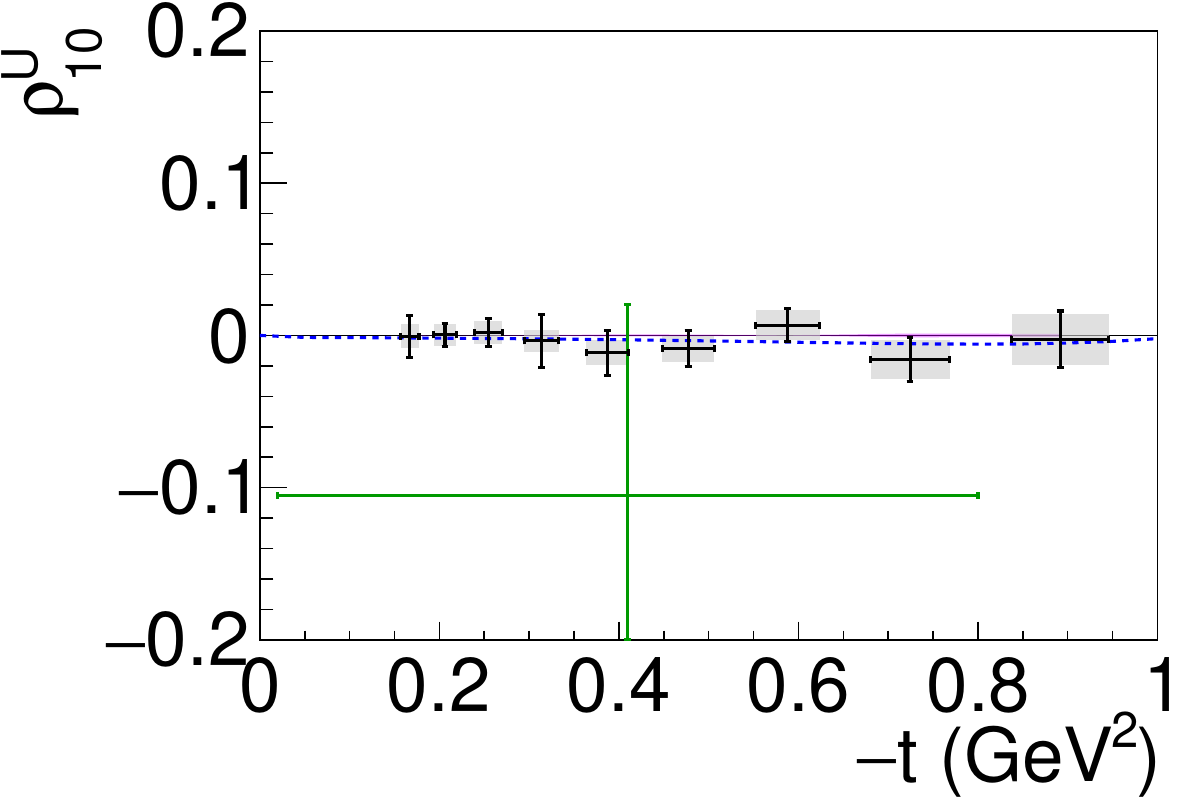}
    \includegraphics[width=0.48\columnwidth]{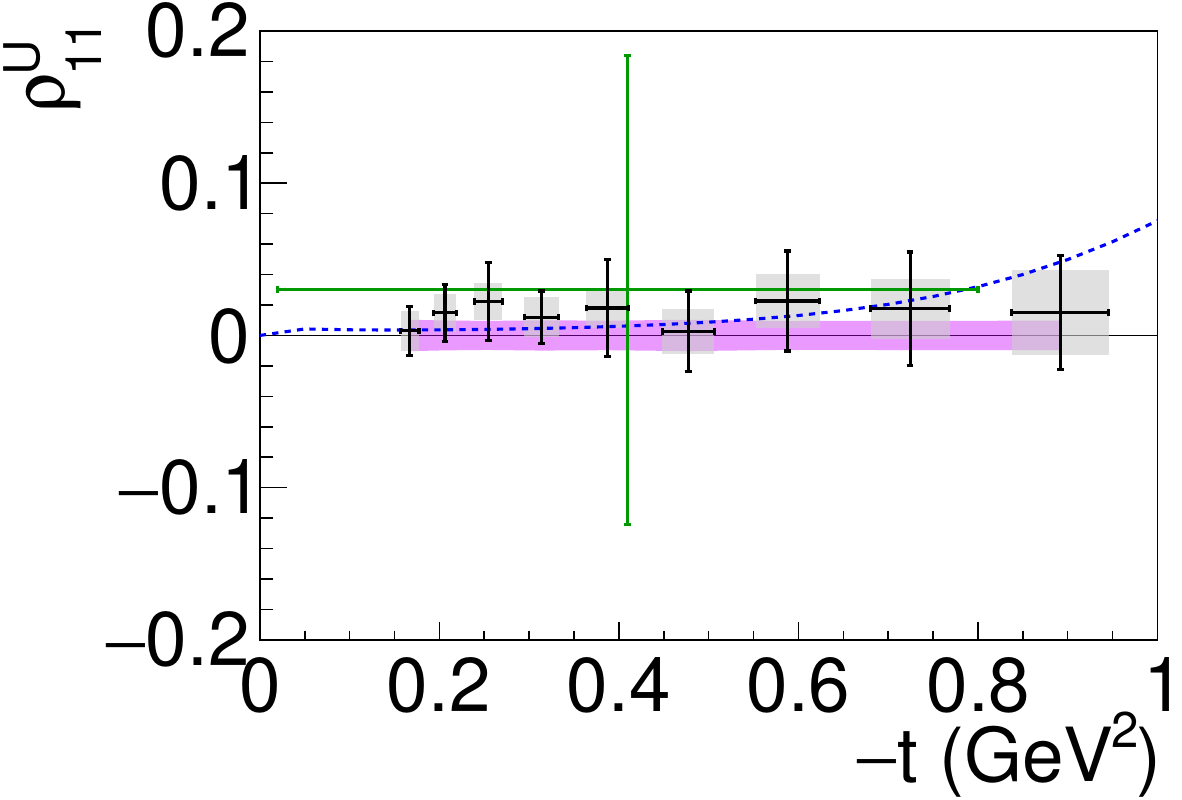}
    \includegraphics[width=0.48\columnwidth]{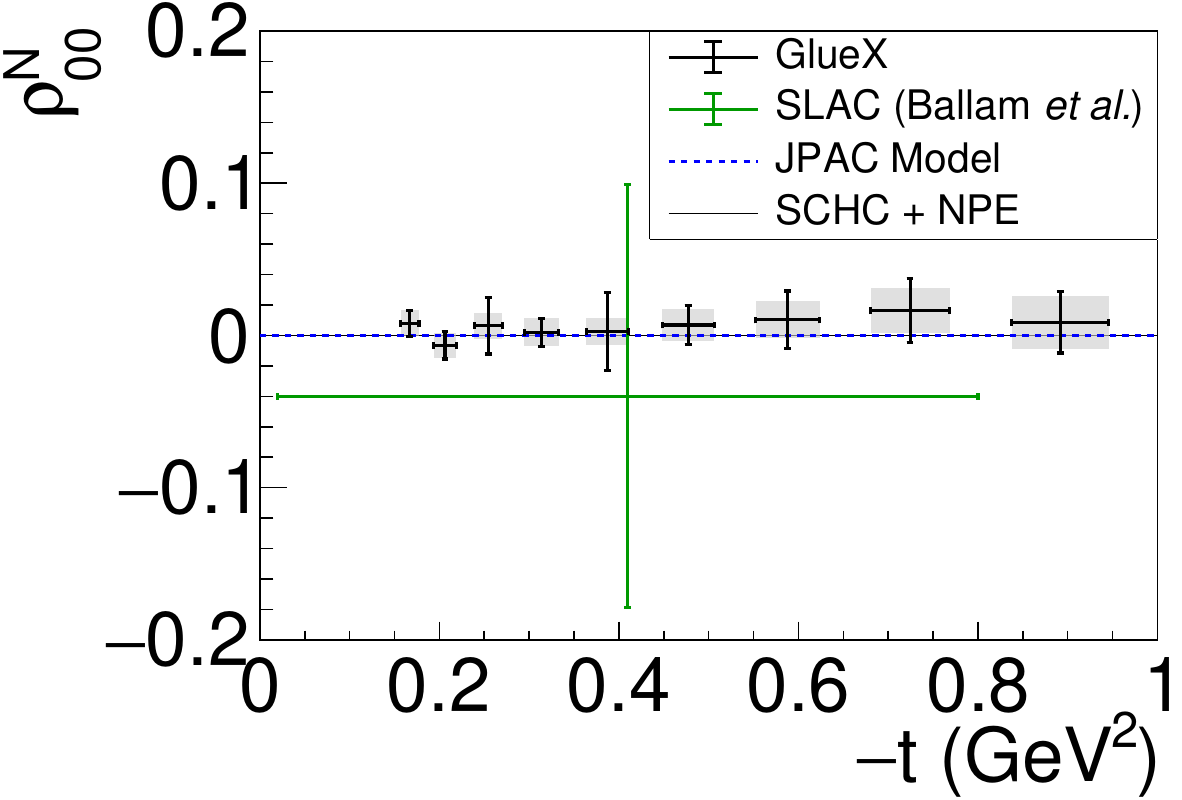}
    \includegraphics[width=0.48\columnwidth]{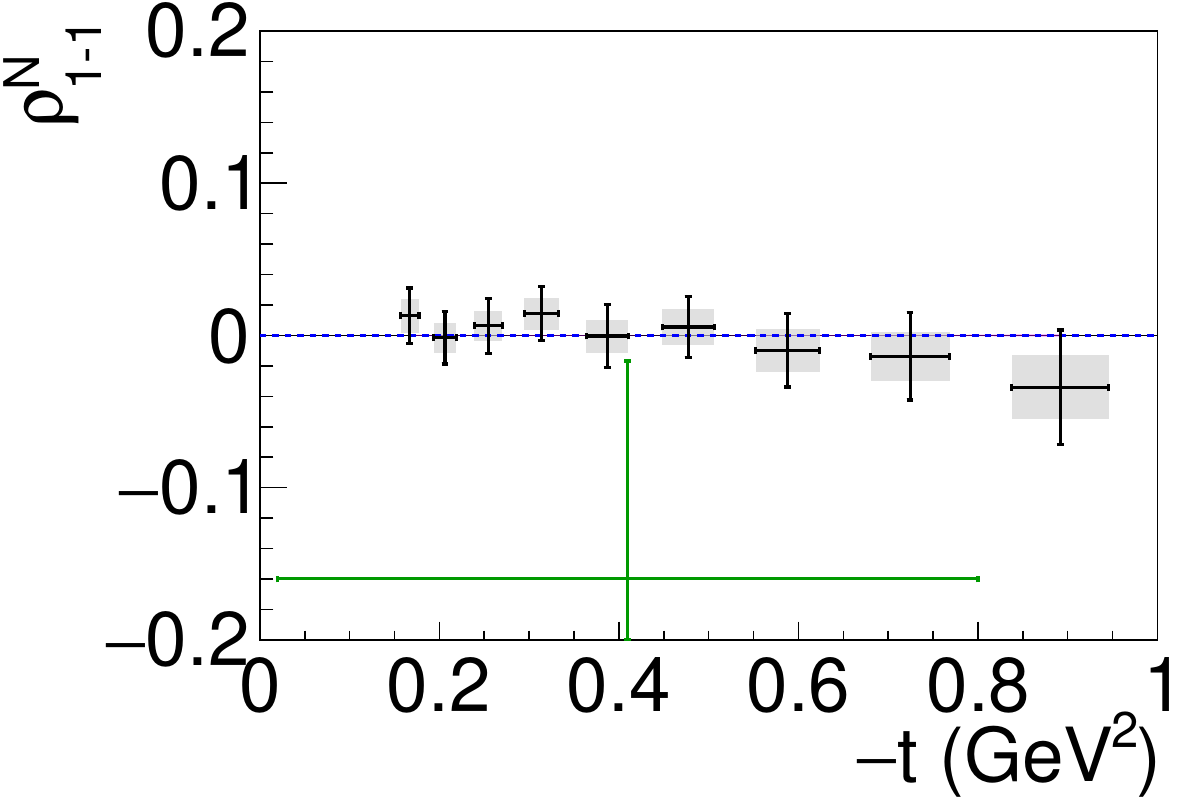}
    \includegraphics[width=0.48\columnwidth]{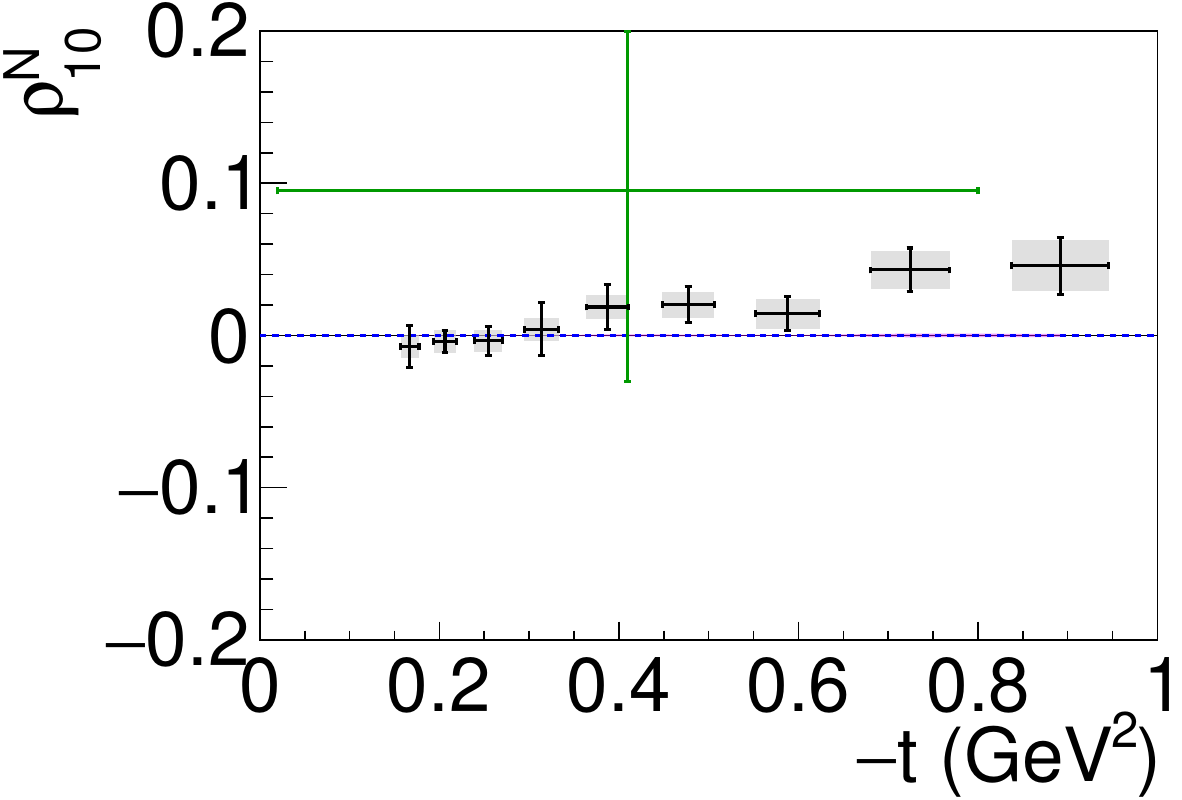}
    \includegraphics[width=0.48\columnwidth]{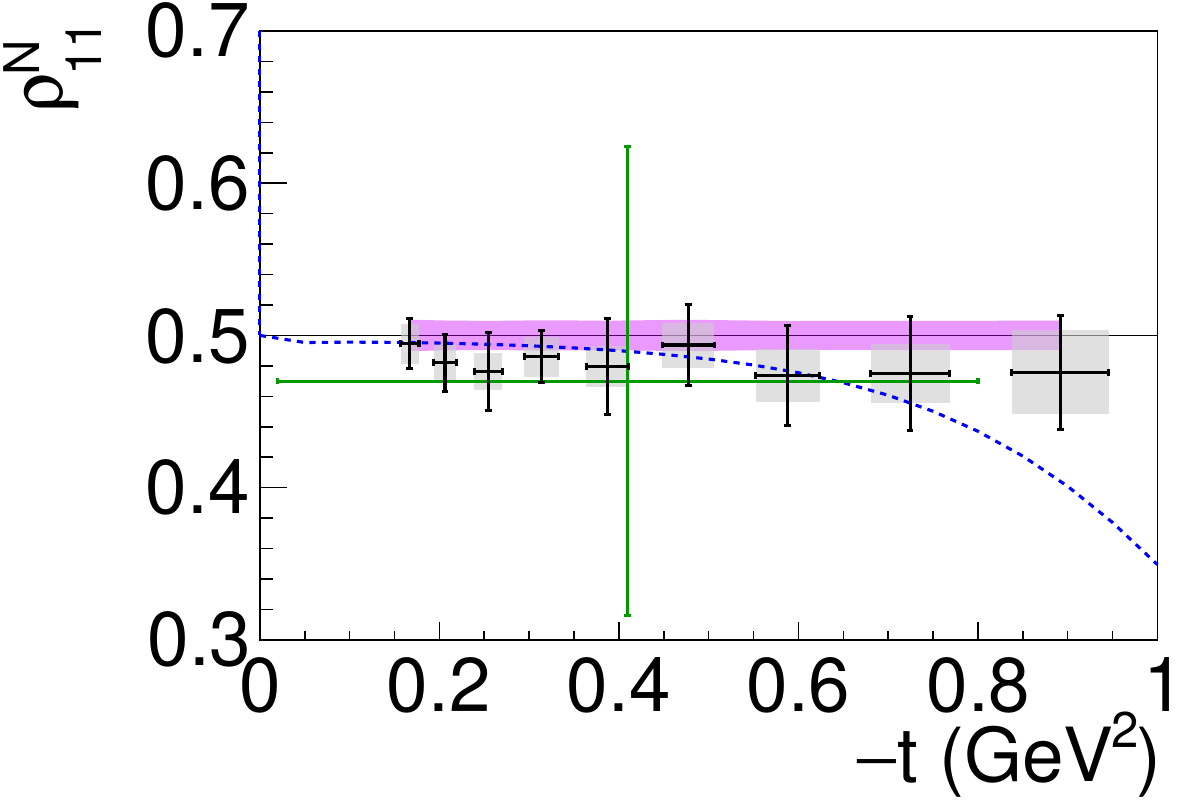}
    \caption{Spin-density matrix elements for $\phi(1020)$ mesons produced by a linearly polarized photon beam for (top row) unnatural- and (bottom row) natural-parity exchange.  The symbols are the same as described in Fig.~\ref{fig:results}.}
    \label{fig:sdme_unnatural}
    \label{fig:sdme_natural}
\end{figure*}

\begin{figure}[!tb]
    \includegraphics[width=0.9\columnwidth]{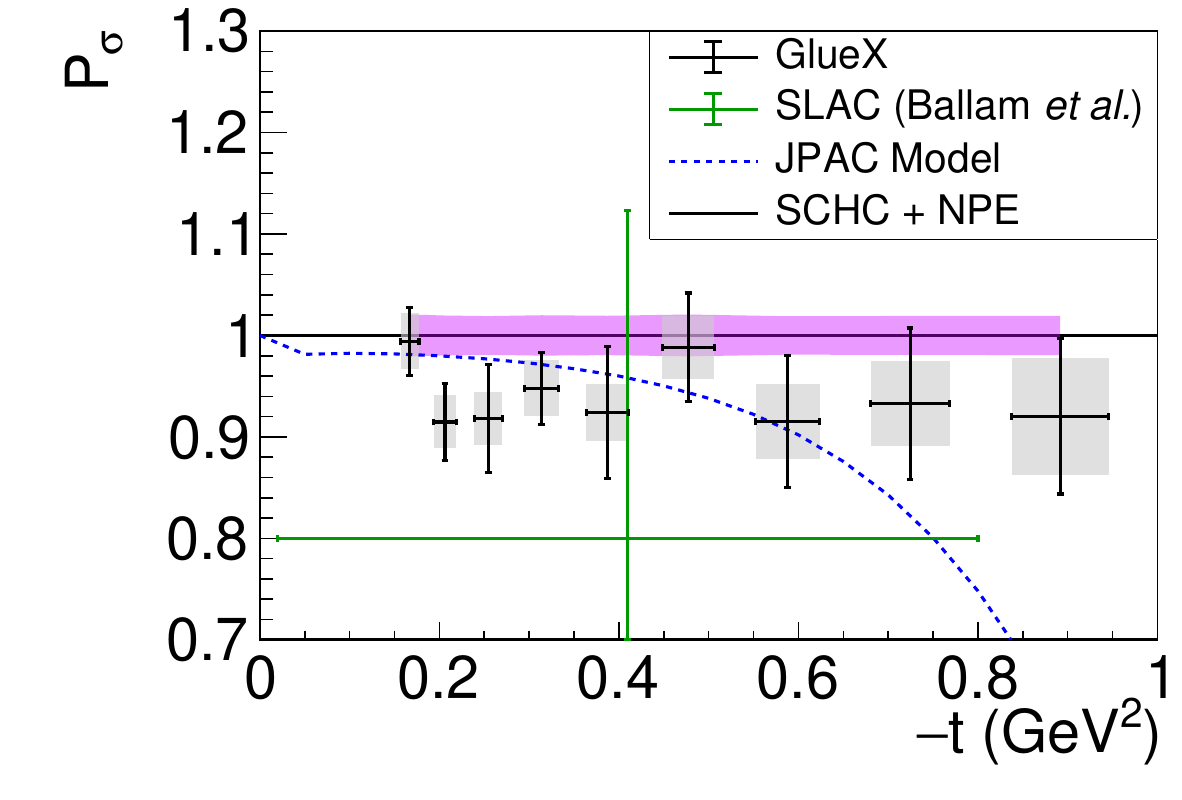}
    \caption{Parity asymmetry for photoproduced $\phi(1020)$.  The symbols are the same as described in Fig.~\ref{fig:results}.}
    \label{fig:psigma}
\end{figure}

\subsection{Parity-Exchange Components}

To better understand the contribution of natural- and unnatural-parity exchanges to the studied process, we separate the spin-density matrix into components $\rho^N_{ij}$ and $\rho^U_{ij}$ that correspond to  natural- ($P=(-1)^J$) or unnatural- ($P = -(-1)^J$) parity exchanges in the $t$-channel, respectively.
There are eight $\rho^{N,U}_{ij}$ components which are given by~\cite{schilling}
\begin{equation}
    \begin{split}
        \rho^{N,U}_{ij} = \frac{1}{2}\left[ \rho^{0}_{ij} \mp (-1)^{-j}\rho^{1}_{i-j} \right],
    \end{split}
\end{equation}
where the $\rho^k_{ij}$ are either measured from the fit to Eq.~4 
or determined from the SDME relations given in Ref.~\cite{schilling}, e.g. $\rho^0_{11}= 0.5\,(1-\rho^0_{00})$.
Our measurements of $\rho^U_{ij}$, shown in the upper row of Fig.~\ref{fig:sdme_unnatural}, are consistent with zero across the analyzed $t$ range.
While the JPAC model predicts a larger unnatural-exchange contribution for $-t\gtrsim0.7$~GeV$^2$ than is consistent with our measurements of $\rho^U_{00}$, a small unnatural-exchange contribution consistent with experimental uncertainties cannot be ruled out.
Our measurements of $\rho^N_{ij}$, shown in the lower row of Fig.~\ref{fig:sdme_natural}, show that $\rho^N_{11}$ is flat in $-t$ and is consistent with 0.5 as expected from Pomeron exchange, and is also consistent with the JPAC model for $-t\lesssim0.7$~GeV$^2$. 
The trend away from zero for  $-t\gtrsim0.7$~GeV$^2$ in $\rho^N_{10}$ and $\rho^N_{1-1}$ suggests a small additional contribution, such as Pomeron couplings that do not conserve helicity.

To leading order, the asymmetry between natural- and unnatural-parity exchange contributions can be expressed in terms of the parity asymmetry $P_\sigma$~\cite{schilling},
\begin{equation}
    P_\sigma = \frac{\sigma^N - \sigma^U}{\sigma^N + \sigma^U} = 2\rho^1_{1-1} - \rho^{1}_{00},
    \label{eq:pSigma}
\end{equation}
where $\sigma^N,\sigma^U$ are the photoproduction cross sections for natural- and unnatural-parity exchange, respectively.  
The parity asymmetry is a quantity normalized between $-1$ and 1 which quantifies the asymmetry of natural- and unnatural-parity exchanges.  A parity asymmetry close to one indicates that natural parity exchange dominates.

The measured parity asymmetry shown in Fig.~\ref{fig:psigma} is independent of $t$ and consistent with natural parity exchange, although systematically less than $+1$ across the analyzed $t$ range.  
These measurements can be compared to the measurement at SLAC of $0.80\pm0.32$ at $E_\gamma=9.3$~GeV~\cite{vector-slac}, 
and by the Omega Photon Collaboration of $P_\sigma = 0.94 \pm 0.34$ at $E_\gamma=20-40$~GeV\cite{phi-omega}, indicating the dominance of natural parity exchange over a large range of photon energies.

Our measurement of the parity asymmetry is consistent with the JPAC model for $-t\lesssim0.7$~GeV$^2$ where Pomeron exchange is dominant.  The $-t$ dependence of the parity asymmetry in the JPAC model occurs because 
 the strength of the natural-parity Pomeron exchange is suppressed as $-t$ increases, and therefore the contributions of the $\pi$ and $\eta$ exchanges become more important, though they are poorly constrained by previous measurements.
 The small systematic deviations of $P_\sigma$ from pure natural-parity exchange suggest either a small, nearly constant contribution from unnatural exchanges, or a small background component underneath the $\phi(1020)$.  We note that including an isotropic background in our fits results in decreases of $P_\sigma$ of $\approx0.02-0.04$, which are  smaller than the statistical uncertainties of these measurements, but are suggestive of the background hypothesis.

\begin{figure}[!tb]
    \centering
    \includegraphics[width=0.9\columnwidth]{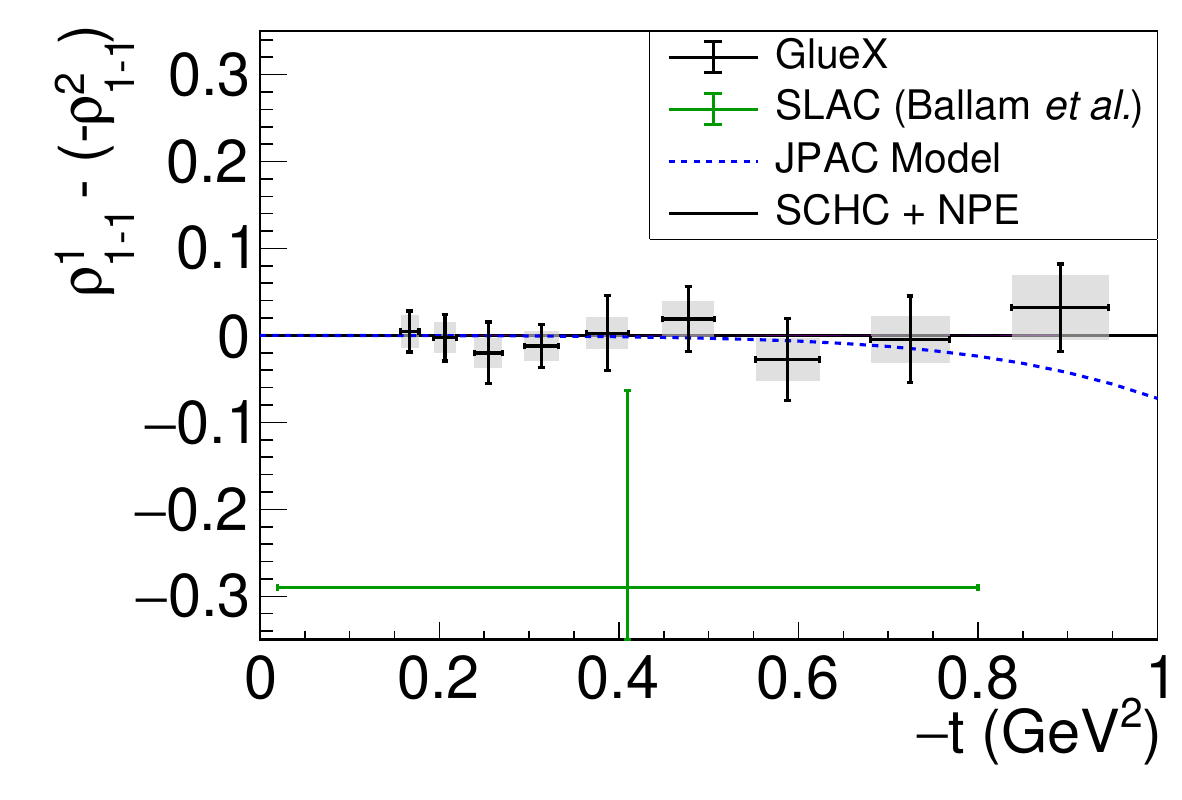}
    \includegraphics[width=0.9\columnwidth]{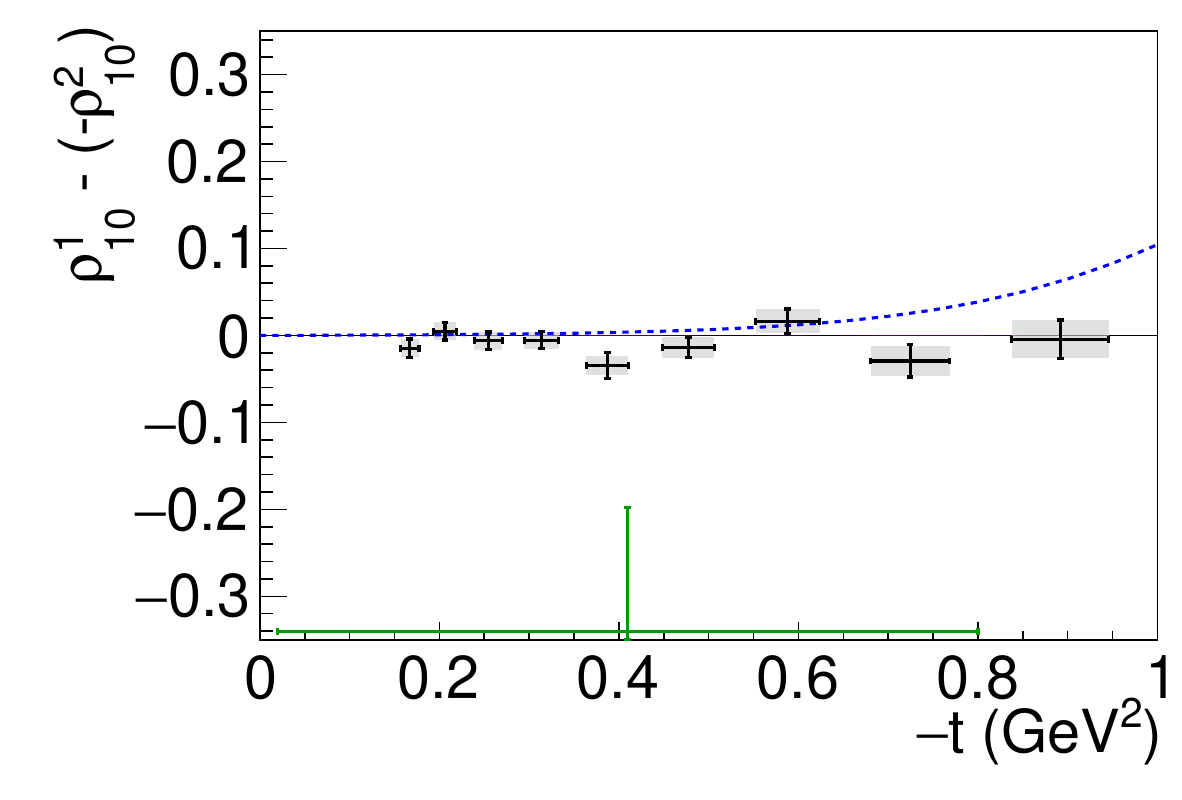}
    \includegraphics[width=0.9\columnwidth]{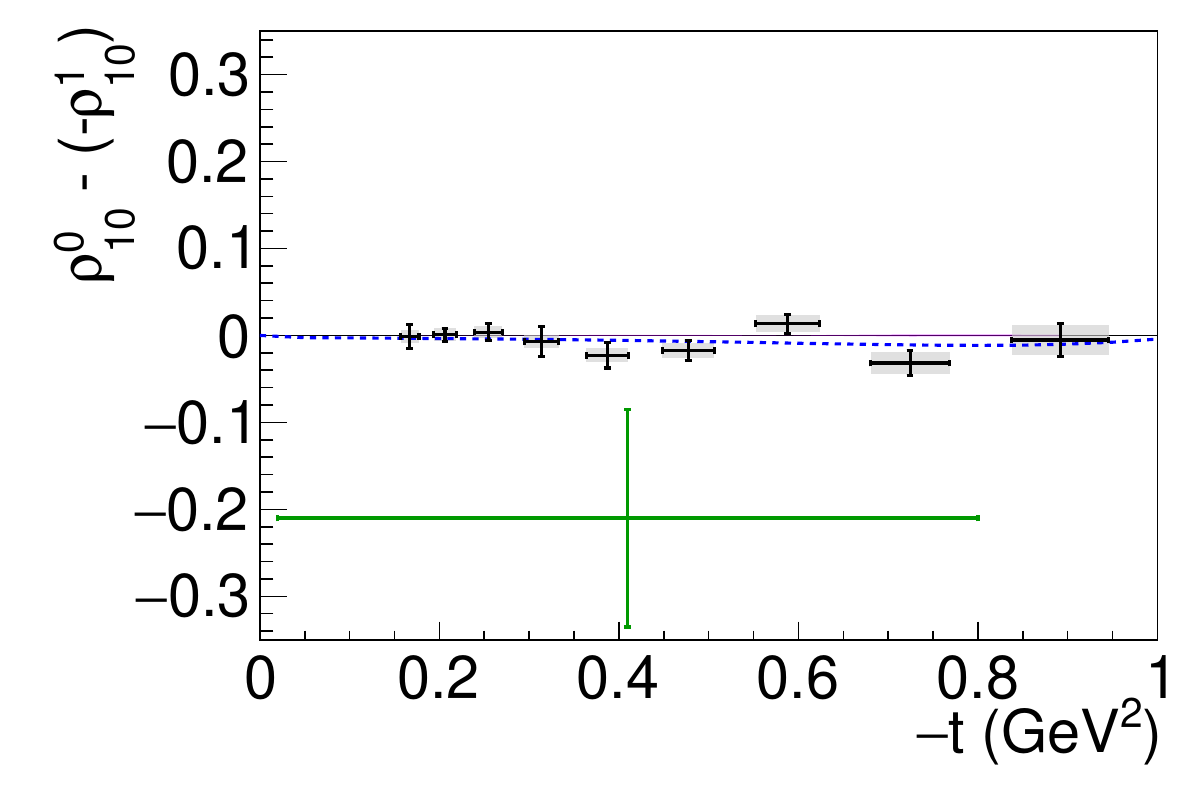}
    \caption{Spin-density matrix element differences, as discussed in the text.  The symbols are the same as described in Fig.~\ref{fig:results}.} 
    \label{fig:sdme_relations}
\end{figure}

\subsection{Relation between SDMEs and Helicity Amplitudes}

As shown in Ref.~\cite{gluex-rho}, neglecting helicity double-flip photoproduction amplitudes that connect the photon and vector-meson helicities which differ by two units leads to the following relations:
\begin{align}
    \rho^1_{1-1} & = -\mathrm{Im}(\rho^2_{1-1}) \\
    \mathrm{Re}(\rho^1_{10}) & = -\mathrm{Im}(\rho^2_{10}) \\
    \mathrm{Re}(\rho^0_{10}) & = \pm \mathrm{Re}(\rho^1_{10}).
    \label{eq:sdme_relation3}
\end{align}
If these relations hold, i.e. if the difference between both SDMEs is zero, helicity double-flip amplitudes are negligible.
Fig.~\ref{fig:sdme_relations} shows each SDME difference as a function of $-t$.
All are consistent with zero across the analyzed $t$ range indicating that contributions from helicity double-flip amplitudes are negligible.  This indicates that $\phi(1020)$ photoproduction is dominated by a single process or by several processes whose amplitudes share the same sign, and is consistent with the expectations of the JPAC model~\cite{Mathieu:2018xyc}.
Concerning Eq.~\ref{eq:sdme_relation3}, we find that $\mathrm{Re}(\rho^0_{10}) = -\mathrm{Re}(\rho^1_{10})$, which is also the case for $\rho(770)$ photoproduction~\cite{gluex-rho}.

\section{Conclusions}

We have measured the spin-density matrix elements of $\phi(1020)$ mesons produced in the scattering of a linearly
polarized photon beam with energy $E_\gamma = 8.2-8.8$~GeV on a proton target using the decay $\phi(1020)\to K_S^0K_L^0$.
These data allow for the first study of the momentum transfer dependence of the SDMEs at these energies, and confirm the dominance of Pomeron exchange in the Regge-theory description of this reaction up to $-t=1.0$~GeV$^2$.
We also find that the contributions from unnatural-parity exchanges are small, and that contributions from helicity-double flip amplitudes are negligibly small.
The precision of our measurements can be improved by including data from the $\phi(1020) \to K^+K^-$ decay and using the data collected during the second GlueX data-collection campaign, which is expected to provide a factor of three more data.
SDMEs may also be measured at larger values of $-t$, although the JPAC model is not expected to describe this reaction beyond $-t\approx1.0$~GeV$^2$. 
Further studies of the reaction $\gamma p \to K_S^0K_L^0 p$ at larger $K_S^0K_L^0$ invariant mass can be expected to provide more insights into excited vector mesons.

\section{Acknowledgments}

We would like to acknowledge the outstanding efforts of the staff of the Accelerator and the Physics Divisions at Jefferson Lab that made the experiment possible. This work was supported in part by the U.S. Department of Energy, the U.S. National Science Foundation, the German Research Foundation, GSI Helmholtzzentrum f\"ur Schwerionenforschung GmbH, the Natural Sciences and Engineering Research Council of Canada, the Russian Foundation for Basic Research, the UK Science and Technology Facilities Council, the National Natural Science Foundation of China and the China Scholarship Council. This material is based upon work supported by the U.S. Department of Energy, Office of Science, Office of Nuclear Physics under contract DE-AC05-06OR23177.
This research used resources of the National Energy Research Scientific Computing Center (NERSC), a U.S. Department of Energy Office of Science User Facility operated under Contract No. DE-AC02-05CH11231. This work used the Extreme Science and Engineering Discovery Environment (XSEDE), which is supported by National Science Foundation grant number ACI-1548562. Specifically, it used the Bridges system, which is supported by NSF award number ACI-1445606, at the Pittsburgh Supercomputing Center (PSC).

\appendix
\section{Numerical Results}

\begin{table*}[!tb]
    \centering
    \caption{Spin-density matrix elements of $\phi(1020)$ mesons produced by a linearly polarized photon beam in the helicity system.  For each bin of $-t$, the limits of the bin range are given, along with the average $-\bar t$ and root-mean-square deviation $-t_\text{RMS}$ of all events that fall within the bin. Each $\rho_{ij}^k$ is shown in units of $\times10^{-3}$.} 
    {\normalsize
    \begin{tabular}{ccccccccccccc}
    \hline\hline
    $-t_\text{min}$ & $-t_\text{max}$ & $-\bar t$ & $-t_\text{RMS}$ & $\rho_{00}^{0}$ & $\rho_{10}^{0}$ & $\rho_{1-1}^{0}$ & $\rho_{11}^{1}$ & $\rho_{00}^{1}$ & $\rho_{10}^{1}$ & $\rho_{1-1}^{1}$ & $\rho_{10}^{2}$ & $\rho_{1-1}^{2}$ \\
    (GeV$^2$) &  (GeV$^2$) &  (GeV$^2$) &  (GeV$^2$) & $\times10^{-3}$ & $\times10^{-3}$ & $\times10^{-3}$ & $\times10^{-3}$ & $\times10^{-3}$ & $\times10^{-3}$ & $\times10^{-3}$ & $\times10^{-3}$ & $\times10^{-3}$ \\  
    \hline
    0.150 & 0.185 & 0.167 &0.010 & $4.9$ & $-7.8$ & $9.4$ & $16.3$ & $-11.0$ & $6.8$ & $491.6$ & $-21.4$ & $-487.1$ \\
    &&&& $\pm2.1$ & $\pm1.8$ & $\pm3.0$ & $\pm10.6$ & $\pm8.2$ & $\pm7.5$ & $\pm13.1$ & $\pm7.2$ & $\pm13.7$ \\
    &&&& $\pm0.4$ & $\pm11.4$ & $\pm14.6$ & $\pm0.3$ & $\pm0.4$ & $\pm0.8$ & $\pm9.6$ & $\pm1.1$ & $\pm10.3$ \\
    0.185 & 0.229 & 0.206 &0.013 & $6.5$ & $-3.6$ & $7.6$ & $-10.8$ & $19.6$ & $4.3$ & $467.2$ & $0.5$ & $-469.8$ \\
    &&&& $\pm2.1$ & $\pm1.8$ & $\pm3.1$ & $\pm9.5$ & $\pm7.8$ & $\pm7.1$ & $\pm12.5$ & $\pm7.1$ & $\pm12.3$ \\
    &&&& $\pm3.0$ & $\pm0.7$ & $\pm14.0$ & $\pm0.7$ & $\pm2.5$ & $\pm1.4$ & $\pm14.0$ & $\pm0.1$ & $\pm14.5$ \\
    0.229 & 0.282 & 0.254 &0.015 & $3.0$ & $-1.6$ & $10.8$ & $2.0$ & $-9.8$ & $5.5$ & $454.0$ & $-11.6$ & $-474.2$ \\
    &&&& $\pm2.1$ & $\pm1.7$ & $\pm3.1$ & $\pm9.3$ & $\pm8.3$ & $\pm7.1$ & $\pm12.3$ & $\pm7.2$ & $\pm12.4$ \\
    &&&& $\pm8.0$ & $\pm5.9$ & $\pm14.1$ & $\pm5.8$ & $\pm14.5$ & $\pm0.2$ & $\pm22.1$ & $\pm0.9$ & $\pm21.7$ \\
    0.282 & 0.349 & 0.314 &0.019 & $4.0$ & $0.4$ & $16.4$ & $12.3$ & $-0.1$ & $-7.6$ & $473.9$ & $2.1$ & $-486.1$ \\
    &&&& $\pm2.0$ & $\pm1.7$ & $\pm3.1$ & $\pm10.1$ & $\pm9.0$ & $\pm7.2$ & $\pm13.1$ & $\pm6.8$ & $\pm12.0$ \\
    &&&& $\pm0.7$ & $\pm15.8$ & $\pm14.4$ & $\pm1.5$ & $\pm0.3$ & $\pm0.1$ & $\pm11.0$ & $\pm0.2$ & $\pm12.8$ \\
    0.349 & 0.430 & 0.387 &0.024 & $4.6$ & $7.3$ & $8.1$ & $-8.9$ & $-0.5$ & $-30.0$ & $461.7$ & $-4.5$ & $-459.0$ \\
    &&&& $\pm2.1$ & $\pm1.8$ & $\pm3.1$ & $\pm10.4$ & $\pm8.5$ & $\pm7.9$ & $\pm13.4$ & $\pm8.0$ & $\pm13.1$ \\
    &&&& $\pm14.9$ & $\pm8.5$ & $\pm15.3$ & $\pm8.7$ & $\pm19.1$ & $\pm8.9$ & $\pm27.7$ & $\pm4.4$ & $\pm27.5$ \\
    0.430 & 0.531 & 0.477 &0.029 & $7.1$ & $11.7$ & $11.3$ & $-0.3$ & $-6.6$ & $-28.9$ & $490.9$ & $15.2$ & $-472.0$ \\
    &&&& $\pm2.6$ & $\pm2.1$ & $\pm3.9$ & $\pm11.0$ & $\pm10.3$ & $\pm8.3$ & $\pm14.7$ & $\pm8.2$ & $\pm15.3$ \\
    &&&& $\pm6.8$ & $\pm8.1$ & $\pm16.6$ & $\pm0.8$ & $\pm0.3$ & $\pm1.2$ & $\pm21.6$ & $\pm0.3$ & $\pm21.4$ \\
    0.531 & 0.656 & 0.588 &0.036 & $7.4$ & $21.1$ & $-6.6$ & $-12.9$ & $-13.0$ & $-7.5$ & $451.0$ & $23.8$ & $-478.7$ \\
    &&&& $\pm2.8$ & $\pm2.4$ & $\pm4.1$ & $\pm13.3$ & $\pm11.8$ & $\pm9.6$ & $\pm17.4$ & $\pm10.1$ & $\pm17.2$ \\
    &&&& $\pm14.5$ & $\pm4.7$ & $\pm19.7$ & $\pm0.1$ & $\pm2.1$ & $\pm1.3$ & $\pm26.9$ & $\pm2.4$ & $\pm30.6$ \\
    0.656 & 0.810 & 0.725 &0.044 & $14.9$ & $27.2$ & $-15.7$ & $-11.6$ & $-17.6$ & $-58.9$ & $457.6$ & $29.5$ & $-462.1$ \\
    &&&& $\pm3.6$ & $\pm2.9$ & $\pm5.3$ & $\pm15.3$ & $\pm14.2$ & $\pm12.4$ & $\pm19.5$ & $\pm12.0$ & $\pm19.2$ \\
    &&&& $\pm15.1$ & $\pm1.8$ & $\pm23.8$ & $\pm0.9$ & $\pm1.7$ & $\pm6.5$ & $\pm30.9$ & $\pm1.7$ & $\pm27.9$ \\
    0.810 & 1.000 & 0.892 &0.054 & $18.3$ & $43.1$ & $-31.3$ & $-36.6$ & $1.1$ & $-48.3$ & $460.7$ & $44.0$ & $-429.0$ \\
    &&&& $\pm4.5$ & $\pm3.6$ & $\pm6.8$ & $\pm20.0$ & $\pm16.7$ & $\pm16.4$ & $\pm27.6$ & $\pm14.0$ & $\pm25.0$ \\
    &&&& $\pm8.6$ & $\pm6.7$ & $\pm31.0$ & $\pm0.4$ & $\pm5.7$ & $\pm4.9$ & $\pm25.0$ & $\pm0.5$ & $\pm23.2$ \\
    \hline\hline
    \end{tabular}
    }
    \label{tab:sdme_values}
\end{table*}

Table~\ref{tab:sdme_values} lists the numerical results for the SDMEs and their statistical and systematic uncertainties.
The systematic uncertainties for the polarized SDMEs $\rho^{1,2}$ contain an overall normalization uncertainty of 2.1\% which is correlated for all bins.
A bias on our measurement of $\rho_{00}^0$ was accounted for by subtracting 0.0075 from the $\rho^0_{00}$ values in all $t$ bins, with an associated systematic uncertainty of 0.0005 assigned to this correction.

\end{document}

%% file: authors.tex
\affiliation{Polytechnic Sciences and Mathematics, School of Applied Sciences and Arts, Arizona State University, Tempe, Arizona 85287, USA}
\affiliation{Department of Physics, National and Kapodistrian University of Athens, 15771 Athens, Greece}
\affiliation{Ruhr-Universit\"{a}t-Bochum, Institut f\"{u}r Experimentalphysik, D-44801 Bochum, Germany}
\affiliation{Helmholtz-Institut f\"{u}r Strahlen- und Kernphysik Universit\"{a}t Bonn, D-53115 Bonn, Germany}
\affiliation{Department of Physics, Carnegie Mellon University, Pittsburgh, Pennsylvania 15213, USA}
\affiliation{Department of Physics, The Catholic University of America, Washington, D.C. 20064, USA}
\affiliation{School of Mathematics and Physics, China University of Geosciences, Wuhan 430074, People’s Republic of China}
\affiliation{Department of Physics, University of Connecticut, Storrs, Connecticut 06269, USA}
\affiliation{Department of Physics, Duke University, Durham, North Carolina 27708, USA}
\affiliation{Department of Physics, Florida International University, Miami, Florida 33199, USA}
\affiliation{Department of Physics, Florida State University, Tallahassee, Florida 32306, USA}
\affiliation{Department of Physics, The George Washington University, Washington, D.C. 20052, USA}
\affiliation{Physikalisches Institut, Justus-Liebig-Universit\"{a}t Gie{\ss}en, D-35390 Gie{\ss}en, Germany}
\affiliation{School of Physics and Astronomy, University of Glasgow, Glasgow G12 8QQ, United Kingdom}
\affiliation{GSI Helmholtzzentrum f\"{u}r Schwerionenforschung GmbH, D-64291 Darmstadt, Germany}
\affiliation{Institute of High Energy Physics, Beijing 100049, People's Republic of China}
\affiliation{Department of Physics, Indiana University, Bloomington, Indiana 47405, USA}
\affiliation{National Research Centre Kurchatov Institute, Moscow 123182, Russia}
\affiliation{Department of Physics, Lamar University, Beaumont, Texas 77710, USA}
\affiliation{Department of Physics, University of Massachusetts, Amherst, Massachusetts 01003, USA}
\affiliation{National Research Nuclear University Moscow Engineering Physics Institute, Moscow 115409, Russia}
\affiliation{Department of Physics, Mount Allison University, Sackville, New Brunswick E4L 1E6, Canada}
\affiliation{Department of Physics, Norfolk State University, Norfolk, Virginia 23504, USA}
\affiliation{Department of Physics, North Carolina A\&T State University, Greensboro, North Carolina 27411, USA}
\affiliation{Department of Physics and Physical Oceanography, University of North Carolina at Wilmington, Wilmington, North Carolina 28403, USA}
\affiliation{Department of Physics, Old Dominion University, Norfolk, Virginia 23529, USA}
\affiliation{Department of Physics, University of Regina, Regina, Saskatchewan S4S 0A2, Canada}
\affiliation{Departamento de Física, Universidad T\'ecnica Federico Santa Mar\'ia, Casilla 110-V Valpara\'iso, Chile}
\affiliation{Department of Mathematics, Physics, and Computer Science, Springfield College, Springfield, Massachusetts, 01109, USA}
\affiliation{Thomas Jefferson National Accelerator Facility, Newport News, Virginia 23606, USA}
\affiliation{Laboratory of Particle Physics, Tomsk Polytechnic University, 634050 Tomsk, Russia}
\affiliation{Department of Physics, Tomsk State University, 634050 Tomsk, Russia}
\affiliation{Department of Physics and Astronomy, Union College, Schenectady, New York 12308, USA}
\affiliation{Department of Physics, Virginia Tech, Blacksburg, VA 24061, USA}
\affiliation{Department of Physics, Washington \& Jefferson College, Washington, Pennsylvania 15301, USA}
\affiliation{Department of Physics, William \& Mary, Williamsburg, Virginia 23185, USA}
\affiliation{School of Physics and Technology, Wuhan University, Wuhan, Hubei 430072, People's Republic of China}
\affiliation{A. I. Alikhanyan National Science Laboratory (Yerevan Physics Institute), 0036 Yerevan, Armenia}
\author{F.~Afzal\orcidlink{0000-0001-8063-6719 }} \affiliation{Ruhr-Universit\"{a}t-Bochum, Institut f\"{u}r Experimentalphysik, D-44801 Bochum, Germany}
\author{C.~S.~Akondi\orcidlink{0000-0001-6303-5217}} \affiliation{Department of Physics, Florida State University, Tallahassee, Florida 32306, USA}
\author{M.~Albrecht\orcidlink{0000-0001-6180-4297}} \affiliation{Thomas Jefferson National Accelerator Facility, Newport News, Virginia 23606, USA}
\author{M.~Amaryan\orcidlink{0000-0002-5648-0256}} \affiliation{Department of Physics, Old Dominion University, Norfolk, Virginia 23529, USA}
\author{S.~Arrigo} \affiliation{Department of Physics, William \& Mary, Williamsburg, Virginia 23185, USA}
\author{V.~Arroyave} \affiliation{Department of Physics, Florida International University, Miami, Florida 33199, USA}
\author{A.~Asaturyan\orcidlink{0000-0002-8105-913X}} \affiliation{Thomas Jefferson National Accelerator Facility, Newport News, Virginia 23606, USA}
\author{A.~Austregesilo\orcidlink{0000-0002-9291-4429}} \affiliation{Thomas Jefferson National Accelerator Facility, Newport News, Virginia 23606, USA}
\author{Z.~Baldwin\orcidlink{0000-0002-8534-0922}} \affiliation{Department of Physics, Carnegie Mellon University, Pittsburgh, Pennsylvania 15213, USA}
\author{F.~Barbosa} \affiliation{Thomas Jefferson National Accelerator Facility, Newport News, Virginia 23606, USA}
% ...
\collaboration{The \textsc{GlueX} Collaboration}